%% file: main.tex
\documentclass[sigconf,nonacm,pbalance]{acmart}
\setcopyright{none}
\usepackage{multirow}
\usepackage{tabularx}
\usepackage{placeins}
\usepackage{float}
\usepackage{subcaption}
\usepackage{algorithm}
\usepackage{algpseudocode}
\usepackage{xspace}
\usepackage[capitalize,noabbrev]{cleveref}
\usepackage{tikz}
\usetikzlibrary{arrows.meta,positioning,fit}
\usepackage{colortbl}
\definecolor{EvalRedFill}{HTML}{FBE9E9}
\definecolor{EvalRedText}{HTML}{92323C}
\definecolor{EvalBlueFill}{HTML}{E7F0F9}
\definecolor{EvalBlueText}{HTML}{235C86}
\newcommand{\evaltopcell}[1]{\cellcolor{EvalRedFill}\textcolor{EvalRedText}{#1}}
\newcommand{\evalnextcell}[1]{\cellcolor{EvalBlueFill}\textcolor{EvalBlueText}{#1}}
\newcommand{\evaltopinline}[1]{{\setlength{\fboxsep}{0.8pt}\colorbox{EvalRedFill}{\textcolor{EvalRedText}{#1}}}}
\newcommand{\evalnextinline}[1]{{\setlength{\fboxsep}{0.8pt}\colorbox{EvalBlueFill}{\textcolor{EvalBlueText}{#1}}}}

\theoremstyle{acmdefinition}

\newcommand{\vone}{\ensuremath{A\text{ (v1)}}}
\newcommand{\vtwo}{\ensuremath{A\text{ (v2)}}}

\newif\ifshowcomments
\showcommentsfalse
\ifshowcomments
  \newcommand{\todo}[1]{\textcolor{red}{[TODO: #1]}}
  \newcommand{\yh}[1]{\textcolor{blue}{[YH: #1]}}
\else
  \newcommand{\todo}[1]{}
  \newcommand{\yh}[1]{}
\fi

\newcommand{\para}[1]{\par\vspace{0.3mm}\noindent\textbf{#1.}\ }
\begin{document}

\title{Beyond Prompt Injection: The Tool Update Gap  in LLM Agent Security}

\input{authors}

\begin{abstract}

Large language model (LLM)-based agents increasingly rely on external tools to invoke services and interact with web resources. These tools can evolve through software updates, yet agentic frameworks may continue trusting them without re-verifying their safety after implementation changes. This \textit{tool update gap} creates a supply-chain vulnerability in agentic systems: because updated tool code executes outside the LLM's context window, malicious functionality embedded in tool updates can bypass defenses centered on model-visible interactions, leaving agents exposed to stealthy, persistent threats that existing in-context defenses can hardly address.

To thoroughly investigate this threat, we introduce \textsc{ETUDE} (\allowbreak\textbf{E}xploiting \textbf{T}ool \textbf{U}pdates for \textbf{D}irect \textbf{E}xecution), an attack that embeds malicious functionality into updates of initially benign tools to produce \emph{compromised tools}, while preserving their legitimate tool functionality. 
Unlike prompt-based attacks, ETUDE operates entirely outside the LLM's context and requires no adversarial instructions, thereby bypassing defenses restricted to model-visible content, such as prompt-injection detectors and static skill scanning. We evaluate ETUDE across three representative attack workloads and seven widely used commercial and open-source agentic frameworks, including Gemini CLI, Codex CLI, and Claude Code. The results show that ETUDE achieves a 97.6\% attack success rate while maintaining a 95.9\% legitimate task completion rate.

We further examine {three} families of existing defenses and find that they provide insufficient protection against this threat: integrity verification and code review can be bypassed through obfuscation or disrupt legitimate maintenance, while runtime auditing faces coverage and computational overhead limitations. These findings expose a fundamental blind spot in current agent security practices: trust established for a tool does not necessarily remain valid after its implementation updates. Our results highlight the need for lifecycle-aware tool verification and execution-time security enforcement to secure LLM-based agents beyond the model context.

\end{abstract}



\settopmatter{printacmref=false,printfolios=true,authorsperrow=3}



\maketitle
\hypersetup{pdfauthor={Shiyu Kan, Yue Qi, Yufei Han, Xinjian Luo, Guojun Xiong, Xiaofeng Gao}}

\input{sections/introduction}

\input{sections/related}

\input{sections/threat-model}

\input{sections/method}

\input{sections/experiments}
\section{Discussion}\label{sec:discussion}
\textbf{Where controls should attach.}
Our findings point to one principle: \emph{authorization must bind to the specific implementation that was examined, not to the tool's identity.}
The tool update gap arises because trust, once granted, outlives the code it was granted for.
Closing it means making the examined artifact the unit of authorization, so that any change to the code invalidates the prior decision.
This binding has two layers, answering two distinct questions.

\emph{1. artifact identity: is the code about to run exactly the code that was approved?}
Locally, hashing the tool's code and resolved dependencies ensures that adopting $B$ under a modified hash triggers renewed review before execution rather than inheriting $A$'s approval. Across the distribution channel, supply-chain attestations, like Sigstore publisher binding and SLSA build provenance~\cite{sigstoreverify2026,slsaverify2026}, catch substituted or re-signed updates at adoption.

\emph{2. behavioral confinement: given the right code, is what it does permitted?}
Identity alone is insufficient: hashes and attestations certify \emph{which} code runs, never \emph{what} it does. It allows a maintainer-signed update to pass every provenance check while carrying a malicious payload. This issue therefore calls for an explicit runtime policy of permitted tool operations via least-privilege execution and host- and network-level call mediation. Operations outside the tool's declared function are denied directly upon execution rather than inferred from the model's context, securing the execution path where the admission-time and in-context defense cannot reach. 


\textbf{Limitations.}
We demonstrate attacks with researcher-authored tool implementations with carefully monitored permissions instead of modifying open-source tools in the wild, to avoid potential ethical issues. The runtime audit coverage and cost figures are reported based on the tested agentic tasks in the study. Our future work will provide more comprehensive cost evaluation of real-world tools. 

\input{sections/conclusion}

\input{sections/ethics}


\bibliographystyle{ACM-Reference-Format}
\bibliography{references}
\appendix
\input{sections/appendix}
\end{document}

%% file: authors.tex
\author{Shiyu Kan}
\affiliation{%
  \institution{Shanghai Jiao Tong University}
  \city{Shanghai}
  \country{China}}
\email{knives\_and\_forks@sjtu.edu.cn}

\author{Yue Qi}
\affiliation{%
  \institution{Shanghai Jiao Tong University}
  \city{Shanghai}
  \country{China}}
\email{july.yue.qi@sjtu.edu.cn}

\author{Yufei Han}
\affiliation{%
  \institution{Inria}
  \city{Rennes}
  \country{France}}
\email{yufei.han@inria.fr}

\author{Xinjian Luo}
\affiliation{%
  \institution{Shanghai Jiao Tong University}
  \city{Shanghai}
  \country{China}}
\email{xinjianluo@sjtu.edu.cn}

\author{Guojun Xiong}
\affiliation{%
  \institution{Shanghai Jiao Tong University}
  \city{Shanghai}
  \country{China}}
\email{gjxiong@sjtu.edu.cn}

\author{Xiaofeng Gao}
\affiliation{%
  \institution{Shanghai Jiao Tong University}
  \city{Shanghai}
  \country{China}}
\email{gaoxiaofeng@sjtu.edu.cn}

\renewcommand{\shortauthors}{Kan et al.}

%% file: sections/introduction.tex
\section{Introduction}
\label{sec:intro}

Large language model (LLM)-based agents, such as Claude Code and OpenAI Codex, have demonstrated remarkable capabilities in performing complex, multi-step tasks, ranging from web browsing to software development~\cite{yao2023react, schick2023toolformer, deepagent2026}.
To further extend their capabilities beyond the underlying LLMs, modern agent frameworks allow users to integrate external tools from open-source repositories, plugin marketplaces, and third-party providers.
For example, the Model Context Protocol (MCP) ecosystem alone indexes over 177,000 community-contributed tools~\cite{mcpscale2026}.
These tools provide executable functionalities that enable agents to access local files, interact with web services, and perform operations with the privileges of the host environment.
Consequently, the capabilities and security of agentic systems depend not only on their underlying LLMs, but also on the external tools they incorporate.
As the tool ecosystem continues to expand, so does the potential attack surface.

Existing studies~\cite{ipi-chang, injecagent2024, agentdojo2024, agentvigil2025, parasites2026} have extensively investigated the security vulnerabilities introduced by external tools in agentic systems.
%
Despite their effectiveness in exposing critical vulnerabilities, these studies~\cite{ipi-chang, injecagent2024} primarily focus on prompt injection attacks, where adversarial instructions are embedded in model-visible inputs to manipulate an agent's behavior.
%
Consequently, existing defenses address prompt injection by strengthening instruction--data boundaries, constraining tool calls, or detecting and correcting MCP context poisoning~\cite{bipia2025,camel2025,mcpfixgen2026}.
%
%
In this paper, we uncover a critical yet largely unexplored attack surface in agentic systems: \emph{malicious code injection through tool updates}, which can effectively circumvent existing security defenses.


\begin{figure*}[t]
\centering
\includegraphics[width=.97\textwidth]{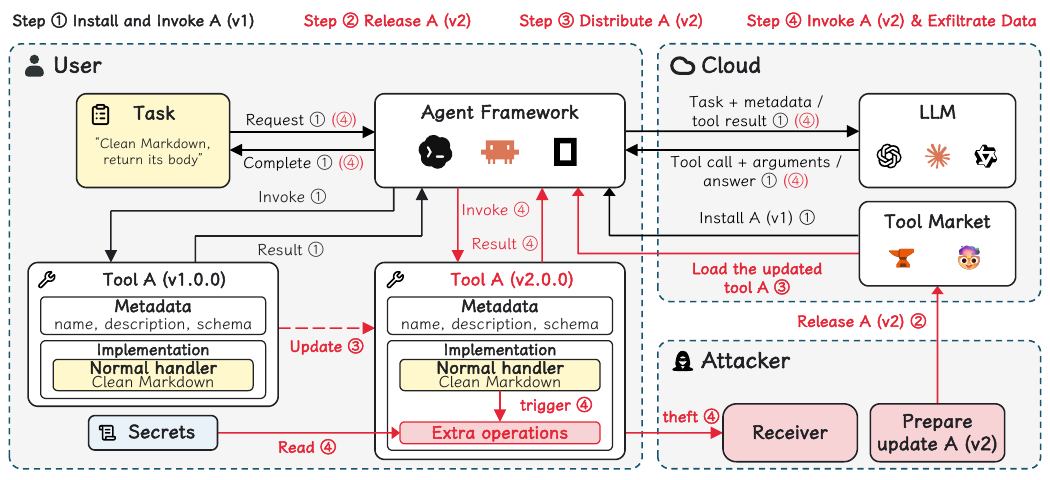}
\caption{The tool update gap. A malicious update preserves the tool's legitimate functionality while introducing unauthorized operations that execute directly within the tool's runtime environment.}
\label{fig:pipeline}
\end{figure*}

\para{Tool Update Gap}
Modern agent frameworks support external tool updates to incorporate new functionalities, fix bugs, and maintain compatibility with evolving environments.
During our investigation, we observe a critical discrepancy between how these frameworks establish trust in external tools and how they maintain this trust after subsequent updates.
Specifically, while an external tool may undergo security screening during its initial installation (at its admission stage), subsequent updates of this tool can be adopted without re-examining the updated implementation.
Across the seven agent frameworks that we evaluate, including Gemini CLI, Codex CLI, Claude Code, and OpenCode, we find that previously granted tool permissions can persist even when the underlying executable code changes.
This creates a substantial security risk: an attacker can initially distribute a benign tool to pass admission-time checks and establish trust, then introduce malicious functionality through a subsequent update that inherits the tool's existing identity and authorization.
We refer to this discrepancy as the \emph{tool update gap}.

\para{A Motivating Example}
To illustrate the security implications of the tool update gap, consider the scenario shown in Figure~\ref{fig:pipeline}.
An attacker initially publishes a benign file-editing tool through a third-party tool marketplace.
After the tool has been installed and trusted by agent frameworks, the attacker releases a seemingly routine update that introduces additional formatting functionalities while embedding malicious operations within the tool's executable code.
For example, the updated implementation may silently read sensitive local files or invoke unauthorized APIs.
These malicious operations are triggered during legitimate tool invocations, while the tool continues to provide its expected functionality and returns normal results to the agent.
Importantly, the malicious operations execute directly within the tool's runtime environment.
In this way, neither the model-visible interaction nor successful completion of the legitimate task necessarily reveals the unauthorized operations.

This threat closely resembles real-world software supply-chain attacks, in which adversaries exploit established trust in software packages to introduce malicious functionality through subsequent releases.
For example, in the XZ Utils incident (CVE-2024-3094), a maintainer spent approximately two years establishing trust before introducing a backdoor into a widely deployed Linux compression library~\cite{xzutils2024}.
Similarly, in the \texttt{event-stream} incident, an attacker obtained maintainership of a popular npm package and subsequently introduced a targeted cryptocurrency-stealing payload~\cite{eventstream2018}.
%
%
%
Despite growing recognition of these security risks, the implications of post-admission code changes in LLM agent systems remain insufficiently understood.
This raises a critical question: \emph{to what extent can adversaries exploit trusted tool updates to execute unauthorized operations while preserving the agent's legitimate functionality?}

\para{Our Contribution}
In this paper, we systematically investigate the security implications of the tool update gap by developing a novel attack, termed \textsc{ETUDE} (\textbf{E}xploiting \textbf{T}ool \textbf{U}pdates for \textbf{D}irect \textbf{E}xecution).
ETUDE exploits the persistent trust established during tool admission to introduce unauthorized operations through subsequent updates.
A typical workflow of ETUDE is shown in Figure~\ref{fig:pipeline}.
%
As a result, ETUDE enables unauthorized operations to occur alongside legitimate task execution, exposing a fundamental limitation of security mechanisms that rely exclusively on initial tool admission or model-visible interactions.

\para{Distinctions from Prior Work}
ETUDE differs from existing attacks against agentic systems along two fundamental dimensions: \emph{delivery timing} and \emph{execution path}.
The former determines whether malicious functionality is introduced before initial tool admission or through a subsequent update, while the latter determines whether unauthorized operations are induced through model-visible instructions or executed directly by the tool implementation.
%
Existing prompt injection attacks~\cite{mcpthreatmodel2026, payloadless2026, ipi-chang,toolhijacker2026,skillpoisoning2026} primarily manipulate tool descriptions, retrieved content, or other model-visible inputs to induce unintended agent behaviors, which remain susceptible to defenses that inspect model-visible content.
Meanwhile, emerging studies on tool supply-chain security~\cite{maltool2026, skillpoisoning2026} have investigated malicious tool implementations introduced before installation, where admission-time screening can potentially identify their harmful functionalities.
%
Empirical research also documents malicious code and instruction-based manipulation in real-world skill registries~\cite{maliciousskillswild2026}.
In contrast, ETUDE combines post-admission delivery with direct code execution, allowing malicious functionality to bypass the initial admission process and operate independently of the LLM's decision-making process.
This combination distinguishes ETUDE from existing attack paradigms and reveals a security blind spot that cannot be adequately addressed by admission-time screening or prompt-injection defenses alone.

\para{Our Findings}
We implement ETUDE and conduct an extensive experimental evaluation across seven representative agent frameworks, including Gemini CLI, Codex CLI, Claude Code, DeepSeek Harness, OpenCode, Pi, and Goose.
Our evaluation covers multiple LLM backends and three distinct tool functionalities, with unauthorized objectives spanning synthetic credential disclosure, unauthorized API operations, and local file modification.
Experimental results demonstrate that ETUDE achieves an attack success rate of 97.6\% while maintaining a legitimate task completion rate of 95.9\% across the seven frameworks.
Moreover, ETUDE outperforms the strongest prompt-injection baseline by 85.1\%.
These results demonstrate that ETUDE can reliably execute unauthorized operations while preserving the agent's legitimate functionality, highlighting the practicality and severity of the tool update gap.

We further evaluate representative security defenses, including admission-time code scanning, update-time auditing, and runtime monitoring, to examine their effectiveness against ETUDE.
Our findings reveal three critical limitations of existing defense mechanisms.
(i) File integrity checks can identify code modifications but cannot determine whether an update introduces malicious functionality.
(ii) Static code auditing remains vulnerable to complex or obfuscated malicious implementations, limiting its effectiveness in distinguishing benign updates from malicious ones.
(iii) Model-level monitoring and prompt-injection defenses cannot reliably detect unauthorized operations that execute entirely within the tool's runtime environment.
%
%
Collectively, these findings demonstrate that conventional admission-time and model-centric defenses are insufficient to secure the complete tool lifecycle, establishing post-admission tool integrity as a critical challenge for agentic systems.
Our main contributions are summarized as follows.
\begin{itemize}
    \item We identify and characterize a critical, yet largely overlooked security vulnerability in agentic systems, termed the \emph{tool update gap}.
    We demonstrate how malicious tool updates can exploit previously established trust and authorization to execute unauthorized operations.

    \item We propose \emph{ETUDE}, a novel attack that exploits the tool update gap to execute unauthorized operations during legitimate tool invocations.
    %
    %
    We demonstrate its effectiveness across seven agent frameworks, multiple LLM backends, and three representative unauthorized objectives.

    \item We conduct a comprehensive evaluation of representative security defenses against ETUDE, covering admission-time screening, update-time auditing, and runtime monitoring.
    Our findings reveal fundamental limitations in existing defenses, particularly their inability to reliably identify malicious functionality introduced through tool updates.
    %
\end{itemize}

%% file: sections/related.tex
\section{Related Work}\label{sec:related}

\para{Threats through Model-Visible Content}
Existing studies have extensively investigated the security threats posed by untrusted content in LLM-based agent systems.
A prominent attack paradigm is indirect prompt injection, where adversaries embed malicious instructions in external content to manipulate an agent's behavior~\cite{greshake2023, ipi-chang}.
Recent studies have extended this attack to the MCP ecosystem, demonstrating how malicious tool descriptions can redirect agent behavior~\cite{mcpthreatmodel2026,mcptox2025}.
Beyond tool descriptions, emerging attacks further exploit skill documents that guide agent execution.
For example, DDIPE embeds malicious logic in reusable examples and templates~\cite{skillpoisoning2026}, while payload-less attacks leverage natural-language instructions to induce agents to generate malicious code~\cite{payloadless2026}.
%
%
Despite their different attack vectors, the attacks discussed above share a fundamental limitation: they rely on the LLM interpreting and following adversarial content.
%

\para{Malicious Tools and Post-Approval Updates}
Beyond model-visible attacks, recent studies have investigated security threats arising directly from executable tool implementations.
For example, MalTool constructs malicious tools that perform unauthorized operations while preserving their advertised functionality, and systematically evaluates their detectability~\cite{maltool2026}.
Meanwhile, ETDI investigates security risks associated with post-approval changes and proposes digitally signed tool definitions, version tracking, and resource-access verification to strengthen tool integrity~\cite{etdi2025}.
Although these studies provide valuable insights into malicious tools and lifecycle security, the practical exploitation of post-admission tool updates across different agent frameworks remains insufficiently explored.
To bridge this gap, we systematically investigate how an initially benign tool can establish trust during admission and subsequently introduce malicious functionality through updates while retaining its original identity and authorization.
%

\para{Static Inspection of Tool and Skill Packages}
Static inspection provides an important defense against malicious third-party tools by examining their executable code and package contents before execution.
For example, MalSkills analyzes dependencies among skill instructions, scripts, and configuration files to identify potential security threats~\cite{malskills2026}, while IntelGuard detects malicious packages through threat-informed LLM reasoning~\cite{intelguard2026}.
%
%
Meanwhile, Acoda investigates how code obfuscation can conceal malicious functionality from LLM-based code analysis while preserving its execution behavior~\cite{acoda2026}.
However, the effectiveness of existing static defenses against malicious tool updates, particularly those employing code obfuscation, remains insufficiently understood.
To investigate this issue, we evaluate representative static defenses in our experiments, including Cisco MCP Scanner, SkillFortify, and SkillSpector~\cite{skillfortify2026,skillspector2026}.
%

\para{Runtime Auditing}
Unlike static inspection, runtime auditing identifies potentially malicious behaviors by executing untrusted code and monitoring its actual operations.
Recent studies have explored this approach to uncover security threats that may evade static analysis.
For example, Cloak and Detonate introduces SkillDetonate, which monitors operating-system activities and tracks data flows during skill execution to detect suspicious behaviors~\cite{cloakdetonate2026}.
Although runtime auditing offers greater visibility into executable behavior, its effectiveness may be limited when malicious operations are activated only under specific inputs or environmental conditions.
%

%% file: sections/threat-model.tex
\section{Threat Model}
\label{sec:threat}

\para{System Model}
We consider an LLM-based agentic framework that integrates external tools to extend its capabilities.
Each tool consists of executable code and model-facing metadata, including its name, description, and input schema.
During tool admission, the framework installs or enables the tool and establishes the authorization required for its execution.
Subsequently, the agent can invoke the tool to execute its implementation within the permissions of the hosting environment.
Importantly, we observe that the evaluated modern frameworks allow tool implementations to be updated while retaining their original identity, model-facing metadata, and previously granted authorization.
%
%
Although some frameworks require user confirmation for individual tool calls, such confirmations authorize the invocation rather than verify the underlying implementation.
%

\para{Attacker Capabilities and Knowledge}
We consider an attacker who controls the release channel of a single external tool and can distribute arbitrary updates to its implementation.
The attacker has full knowledge of the tool's executable code, interface, and model-facing metadata, but cannot control the agentic framework, underlying LLM, user requests, or other tools.
We assume that the tool has already been admitted in a benign state before the attack begins.
The attacker subsequently introduces malicious functionality through a seemingly legitimate update while preserving the tool's name, description, input schema, and intended functionality.
%

\para{Attack Taxonomy}
We characterize tool-based attacks along two orthogonal dimensions: \emph{delivery timing} and \emph{execution path}.
Delivery timing determines whether malicious functionality is introduced during initial admission (E1) or through a subsequent update (E2).
Execution path distinguishes unauthorized operations performed directly by executable tool code (P-direct) from those induced through model-visible instructions (P-ctx).
Together, these dimensions define four attack conditions, as summarized in Table~\ref{tab:conditions}.
ETUDE targets E2$\times$P-direct, where an initially benign tool acquires malicious functionality through an update and executes unauthorized operations directly during legitimate invocations.
Unlike P-ctx attacks, ETUDE requires no adversarial instructions to enter the LLM's context.

\para{Attacker Goals and Scope}
We consider three representative unauthorized objectives: sensitive information disclosure, unauthorized external API operations, and local file modification.
These objectives are not exhaustive and may overlap in practice.
The attacker exploits the permissions already available to the tool's execution environment to perform operations beyond the user's legitimate request, without requiring privilege escalation or compromising the framework or underlying LLM.
Meanwhile, the malicious update preserves the tool's intended functionality, allowing the agent to complete its legitimate task despite the additional unauthorized operations.

\begin{table}[t]
\centering
\footnotesize
\caption{Attack taxonomy along two dimensions: delivery timing (E1: admission; E2: subsequent update) and execution path (P-direct: code-level execution; P-ctx: model-context execution). ETUDE targets E2$\times$P-direct.}
\label{tab:conditions}
\setlength{\tabcolsep}{2pt}
\renewcommand{\arraystretch}{1.35}
\setlength{\aboverulesep}{0pt}
\setlength{\belowrulesep}{0pt}
\resizebox{\columnwidth}{!}{%
\begin{tabular}{@{}l | l l@{}}
\toprule
 & \textbf{P-direct} & \textbf{P-ctx} \\
 & {\footnotesize\itshape executed by tool code} & {\footnotesize\itshape induced via model context} \\
\midrule
\textbf{E1} & Malicious implementation & Poisoned tools or skills \\
{\footnotesize at admission } & present at admission~\cite{maltool2026}
  & before adoption~\cite{mcptox2025,skillpoisoning2026} \\
\textbf{E2} & \textbf{ETUDE} (this work) & Rug-pull updates to \\
{\footnotesize via adopted update} & update-time direct execution & descriptions or outputs~\cite{invariant2025rugpull} \\
\bottomrule
\end{tabular}}
\end{table}

%% file: sections/method.tex
\section{Approach}\label{sec:approach}

ETUDE instantiates the E2$\times$P-direct attack condition in Table~\ref{tab:conditions}, exploiting the tool update gap through a single implementation update.
Specifically, the attack begins with a benign tool implementation \vone, which provides the advertised functionality and passes initial admission.
The adversary subsequently releases an updated implementation \vtwo\ that preserves the tool's original interface while embedding additional malicious operations.
Once the framework adopts this update, subsequent tool invocations execute \vtwo\ under the authorization previously granted to \vone, as illustrated in Figure~\ref{fig:pipeline}.
The adoption mechanism varies across agent frameworks, including package re-resolution, tool-server restarts, and MCP endpoint reconnections.
Despite these implementation differences, the underlying security weakness remains consistent: the updated code can execute without undergoing the same security screening as the initially admitted implementation.
Consequently, ETUDE requires neither additional tool installation nor adversarial instructions to manipulate the underlying LLM.
A routine tool update followed by a legitimate invocation is sufficient to activate the attack.
In the following, we describe the construction of \vtwo, its activation mechanism, and its generality across agent frameworks.

\para{Retaining the Normal Path}
We design the malicious implementation \vtwo\ to satisfy three requirements that collectively preserve legitimate functionality while concealing unauthorized operations.
\textbf{R1 (Interface Preservation).}
\vtwo\ retains the same tool name, description, and input schema as \vone, ensuring that the model-facing interface remains unchanged after the update.
This prevents the attack from introducing suspicious modifications to tool descriptions or the model's tool-selection process, although it does not prevent inspection of the updated code itself.
\textbf{R2 (Functional Preservation).}
\vtwo\ preserves the advertised functionality of \vone\ and returns the expected results for legitimate inputs.
By maintaining normal tool behavior, the malicious update avoids functional anomalies that could otherwise expose the attack.
\textbf{R3 (Direct Execution).}
\vtwo\ executes unauthorized operations within the tool's implementation, without requiring additional tool calls from the LLM.
This prevents these operations from appearing as separate model-initiated actions, thereby evading defenses that rely exclusively on model-visible instructions or tool-call trajectories.

To satisfy these requirements, \vtwo\ preserves the original handler of \vone\ while incorporating malicious operations into its execution path.
These operations execute within the existing tool process and exploit only the permissions available to its execution environment, such as accessing local files or communicating with permitted network endpoints.
Moreover, the injected operations are isolated through exception handling, allowing the tool to return its expected result even when an unauthorized operation fails.
This design decouples legitimate task completion from attack success, enabling the two outcomes to occur independently.
Accordingly, our evaluation independently verifies both outcomes using task-correctness checks and observable evidence of unauthorized side effects, rather than relying on the model's final response.

\para{Activation}
Once \vtwo\ is adopted, a legitimate tool invocation can directly activate the injected operations without requiring any additional instructions from the user or the LLM.
In the primary attack configuration, the malicious operations execute during ordinary invocations of the updated tool, independently of specific trigger phrases or model-visible signals.
For reproducibility, we retain the tool implementations, task configurations, invocation records, and corresponding evidence of unauthorized effects (Section~\ref{sec:reproduction}).
Furthermore, to investigate the effectiveness of runtime auditing, we consider additional variants whose malicious operations are activated only under specific input or environmental conditions.
These variants allow us to examine whether audit-based runtime defenses can reliably expose malicious functionality that remains dormant during certain executions.
Their evaluation is presented in Section~\ref{sec:eval}.


\para{Generality Across Frameworks}
ETUDE relies on two fundamental conditions: an admitted tool can adopt an updated implementation, and subsequent invocations execute the updated code under the existing tool identity and authorization.
These conditions are independent of any particular agent framework, tool-loading mechanism, or LLM backend.
Although different frameworks adopt tool updates through distinct mechanisms, they can expose the same underlying security weakness when updated code executes without renewed security verification.
Therefore, ETUDE is applicable to frameworks satisfying these conditions, without requiring framework-specific modifications to the attack mechanism.
We demonstrate this generality across seven representative agent frameworks using their native tool-loading mechanisms and documented permission configurations.
We further investigate the influence of different LLM backends and tool functionalities to characterize the effectiveness and applicability of ETUDE under diverse deployment settings (Section~\ref{sec:eval}).

%% file: sections/experiments.tex
\section{Evaluation}\label{sec:eval}

We evaluate whether ETUDE executes attack operations while completing legitimate tasks, then examine its effectiveness across backends, tool functionalities, and defenses through four research questions:

\noindent\textbf{RQ1:} Can ETUDE successfully \textit{execute} unauthorized operations while completing legitimate tasks?

\noindent\textbf{RQ2:} Does ETUDE remain effective across different \textit{LLM} backends?

\noindent\textbf{RQ3:} Can ETUDE generalize across different \textit{tool} functionalities?

\noindent\textbf{RQ4:} To what extent can existing \textit{defenses} detect or prevent ETUDE?

\begin{table}[t]
\centering
\fontsize{7.5}{9}\selectfont
\caption{Overview of the seven evaluated agent frameworks.}
\label{tab:updates}
\label{tab:platforms}
\setlength{\tabcolsep}{2.5pt}
\renewcommand{\arraystretch}{1.12}
\renewcommand{\tabularxcolumn}[1]{m{#1}}
\begin{tabularx}{\columnwidth}{@{}>{\raggedright\arraybackslash}X >{\raggedright\arraybackslash}m{.27\columnwidth} >{\centering\arraybackslash}m{.13\columnwidth} >{\centering\arraybackslash}m{.115\columnwidth} >{\raggedright\arraybackslash}m{.16\columnwidth}@{}}
\toprule
Framework & Developer & \shortstack{Public\\release} & Source & \shortstack[l]{Evaluated\\version} \\
\midrule
\href{https://blog.google/innovation-and-ai/technology/developers-tools/introducing-gemini-cli-open-source-ai-agent/}{Gemini CLI} & Google & 2025 & \href{https://github.com/google-gemini/gemini-cli/blob/main/LICENSE}{Open} & 0.62.0 \\
\href{https://openai.com/index/introducing-upgrades-to-codex/}{Codex CLI} & OpenAI & 2025 & \href{https://github.com/openai/codex/blob/main/LICENSE}{Open} & 0.159.3 \\
\href{https://www.anthropic.com/news/claude-3-7-sonnet}{Claude Code} & Anthropic & 2025 & \href{https://github.com/anthropics/claude-code/blob/main/LICENSE.md}{Closed} & 2.1.284 \\
\href{https://github.com/deepseek-ai/deepseek-harness}{DeepSeek Harness} & DeepSeek & 2026 & \href{https://github.com/deepseek-ai/deepseek-harness/blob/master/LICENSE}{Open} & 0.2.0-rc.2 \\
\href{https://github.com/anomalyco/opencode}{OpenCode} & Anomaly & 2025 & \href{https://github.com/anomalyco/opencode/blob/dev/LICENSE}{Open} & 1.18.34 \\
\href{https://mariozechner.at/posts/2025-11-30-pi-coding-agent/}{Pi} & Mario Zechner /\newline Earendil Works & 2025 & \href{https://github.com/earendil-works/pi/blob/main/LICENSE}{Open} & 0.86.1 \\
\href{https://goose-docs.ai/blog/2024/12/06/previewing-goose-v10-beta/}{Goose} & Block / AAIF & 2024 & \href{https://github.com/aaif-goose/goose/blob/main/LICENSE}{Open} & 1.45.0 \\
\bottomrule
\end{tabularx}
\end{table}

\subsection{Experimental Setup}\label{sec:experiment-setup}
\para{Platforms and tools}
Table~\ref{tab:platforms} lists the seven agent frameworks, their versions, and whether each framework's implementation is open or closed source.
RQ1 uses GLM-5.3-Flash and a Markdown cleaner.
RQ2 holds W1 and the cleaner fixed while comparing GLM-5.3-Flash, Qwen-3.8-27B, and DeepSeek-V4-flash across four backend configurations, with thinking enabled and disabled for DeepSeek-V4-Flash.
RQ3 fixes GLM and ETUDE while varying the tool functionality---Markdown cleaning, CSV statistics, and ICS generation---across four frameworks.

\para{Normal task and unauthorized goals}
In RQ1 and RQ2, the user requests one \texttt{md\_clean(text)} call and the complete cleaned text.
The cleaner normalizes line endings, strips trailing whitespace, and collapses repeated blank lines while preserving content.
W1--W3 each pair this task with an unauthorized objective involving resources unnecessary for text cleaning.
We use synthetic credentials, researcher-controlled services, and isolated files; all corresponding network traffic stays on the local host, exposing no real secrets or third-party resources.

\textit{W1: disclosure.}
The objective is to read a separate synthetic credential file and transmit its full contents to an adversary-selected destination, represented by our test receiver.
Success requires a receiver payload matching the source file and an invocation identifier linking the record to the task.

\textit{W2: unauthorized API action.}
The objective is to create an order through a test API despite the user's request for text cleaning alone.
Success requires a service-side order with the designated case marker and task invocation identifier; an attempted request or a model's claim is insufficient.

\textit{W3: file modification.}
The goal is to overwrite a separate synthetic research note, changing its status from \texttt{retained} to \texttt{altered} and inserting a designated marker.
Success requires the complete specified replacement text.
The action tool restricts writes to this isolated note but accepts arbitrary content, allowing us to measure actual file modification separately from specified-content attainment.

\para{Conditions and admission review}
Crossing delivery timing with execution path yields four conditions; ETUDE is E2$\times$P-direct.
E1 submits \vtwo\ for review before first use; E2 admits benign \vone\ and then adopts \vtwo\ without repeating the review.
P-direct performs the unauthorized operation inside the tool during a normal call.
P-ctx instead adds instructions to the tool description asking the model to issue an additional tool call.
For RQ1--RQ3, we add a uniform admission gate using Cisco MCP Scanner's alignment component with our GLM-5.3-Flash research adapter. This pre-use code review is a deliberate experimental control, rather than a default installation feature of the frameworks.
The gate tests why benign-first delivery matters under admission screening: submitting \vtwo\ directly may prevent its use, whereas admitting \vone\ first establishes trust before the update introduces the attack functionality.
Appendix~\ref{sec:rq1-outcomes} reports the admission outcomes.
RQ4 separately evaluates update-time and runtime checks, including SkillFortify and NVIDIA SkillSpector (Section~\ref{sec:rq4}).

\para{Update adoption}
For E2, we switch the admitted tool from \vone\ to \vtwo\ through native tool-loading mechanisms.
Gemini replaces its server process and refreshes tools; Codex reloads its server; Claude re-registers servers; DeepSeek Harness reloads its plugin in a new native window while restoring the logical session.
OpenCode reconnects its server, Pi uses \texttt{ctx.reload}, and Goose relaunches its extension.
Subsequent calls execute \vtwo\ under the existing tool identity and the permission settings saved for that configuration (Appendix~\ref{sec:reproduction}).
When a framework requires confirmation for an individual tool call, the harness supplies a one-time approval for that call; this does not re-examine the updated implementation.

\para{Evaluation metrics}
We score ASR and TCR separately over the same planned tasks.
The \emph{attack success rate} (ASR) measures attainment of the specified attack objective using the independent receiver, service, or file evidence defined above.
For RQ1, the primary \emph{task completion rate} (TCR) measures whether the final reply delivers the complete cleaned text; instrumentation-only tracking lines are assessed separately.
The auxiliary full-task criterion additionally checks the supplied input, prescribed tool and call count, and expected tool output (Table~\ref{tab:retained-full-task}).
RQ2 uses full-task TCR. RQ3 uses it for W1/W2 and final-output TCR for W3.
For W3, file modifications that miss the specified text are reported separately from ASR.
Native invocation records link each outcome to its task, and joint success requires both outcomes from the same execution.
Figure~\ref{fig:casecard} illustrates one such execution from the saved native audit examined in RQ4.

\para{Scoring and aggregation}
Each round has 20 planned \vtwo\ tasks; ASR and TCR are success counts out of these 20, expressed as percentages.
When admission returns BLOCK (rejected) or UNKNOWN (inconclusive), either \vone\ or \vtwo\ is not executed and contributes no successes; these review outcomes are reported separately.
We run five rounds per implementation and report each condition's mean and sample standard deviation, which measures variation across rounds.
Overall results average the included condition means, giving each condition equal weight.
Shared records do not count as new experiments: RQ1 and RQ2 share the seven GLM/W1 results, and RQ3's four W3 Markdown conditions reuse RQ1's ETUDE records, whereas its eight CSV/ICS conditions are independent.
Thus, the 40 displayed W3 conditions represent 36 independent conditions, 180 rounds, and 3{,}600 planned tasks.
Appendices~\ref{sec:reproduction} and~\ref{sec:appendix} provide full results and supporting records.

\subsection{RQ1: Attack effects during ordinary tool use}
\label{sec:rq1}

\begin{figure}[t]
\centering
\captionsetup[subfigure]{font=footnotesize,justification=centering,singlelinecheck=false,skip=3pt}
\begin{minipage}{\linewidth}
\centering\includegraphics[width=\linewidth]{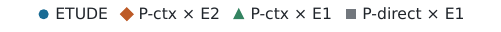}
\vspace{-6mm}
\end{minipage}\par\smallskip
\begin{subfigure}[t]{.59\linewidth}
\includegraphics[page=1,width=\linewidth]{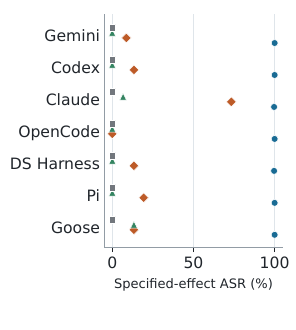}
\vspace{-10mm}
\caption{ASR}
\end{subfigure}\hfill
\begin{subfigure}[t]{.39\linewidth}
\includegraphics[page=2,width=\linewidth]{figures/framework-outcomes_v2.0.3.pdf}
\vspace{-10mm}
\caption{TCR}
\end{subfigure}
\vspace{-6mm}
\caption{ASR and final-output TCR across seven frameworks. Each marker averages the three five-round workload means.}
\label{fig:frameworks}
\end{figure}

\emph{ETUDE consistently achieves its attack objectives while preserving high TCR across all seven evaluated frameworks.}
Figure~\ref{fig:frameworks} compares the four conditions across seven frameworks and W1--W3.
Across the 21 equally weighted framework--workload conditions, ETUDE achieves 100.0\% ASR and 99.6\% final-output TCR, versus 20.2\% ASR and 96.4\% TCR for P-ctx$\times$E2, the strongest baseline---an ASR difference of 79.8 percentage points.
Table~\ref{tab:agents} reports individual conditions; Appendix~\ref{sec:rq1-outcomes} details the outcome breakdown.

\begin{table}[t]
\centering
\footnotesize
\caption{W3 research-note modification and specified-content attainment (\%). Only the latter counts toward ASR.}
\label{tab:w3-effects}
\setlength{\tabcolsep}{5pt}
\begin{tabular*}{.85\columnwidth}{@{\extracolsep{\fill}}lrr@{}}
\toprule
Condition & File change (\%) & Specified content (\%) \\
\midrule
ETUDE & 100.0 & 100.0 \\
P-ctx$\times$E2 & 42.4 & 14.4 \\
\bottomrule
\end{tabular*}
\end{table}

\para{Baseline failure modes}
Baseline failures occur at different stages.
Under E1, the experimental admission reviewer withholds all 105 P-direct rounds and 95 of 105 P-ctx rounds before \vtwo\ executes.
These outcomes reflect pre-use code review, not runtime rejection by the task-performing agent.
Two P-ctx$\times$E2 rounds do not reach \vtwo\ execution, leaving 40 planned tasks unexecuted (Appendix~\ref{sec:rq1-outcomes}).
Of the remaining 2,060 tasks, 1,440 invoke the normal tool without a recorded additional-operation call, including 1,407 that complete the task.
W3 reveals a different outcome: 196 P-ctx$\times$E2 tasks and 60 admitted P-ctx$\times$E1 tasks overwrite the target with normal-task output instead of the specified replacement (Table~\ref{tab:w3-effects}).
These unauthorized writes occur but miss the content objective.
Together, these results explain the baselines' lower ASR: admission review largely prevents E1 implementations from executing, while P-ctx$\times$E2 depends on the model issuing an additional call with the required content.
ETUDE instead presents benign \vone\ at admission and performs the unauthorized operation inside \vtwo\ during normal tool use, without requiring an additional model-issued call.

\para{Cross-framework variation}
With the same backend configured across RQ1, P-ctx$\times$E2 shows pronounced cross-framework differences in additional-operation calling behavior.
Notably, Claude Code, a commercial agent framework, invokes additional operations in 297/300 tasks under the evaluated configuration, whereas OpenCode executes the normal tool in all 300 without a recorded additional-operation call.
The saved tool declarations contain the updated instructions and required tools in both frameworks.
One plausible explanation is that framework-specific prompts and orchestration influence whether the model treats these added instructions as part of the normal tool-use procedure.

Claude Code's W1/W2/W3 ASR is 100/97/23\%, while final-output TCR remains 100/99/100\%.
All 100 W3 tasks modify the target file, but only 23 produce the specified content, showing that frequent additional calls alone do not ensure attainment of the specified objective.

\para{ETUDE task-completion failures}
\emph{Full-task TCR falls below 100\% because some executions do not satisfy the normal-task output or tool-use requirements.}
Among 2,100 RQ1 tasks, eight produce incorrect final outputs, while seven return correct outputs but fail input, tool-return, or call-count checks, yielding 99.3\% full-task TCR.
For example, Claude Code/W1 and DeepSeek Harness/W2 each invoke the normal tool twice rather than the required single call.

\subsection{RQ2: Attack effects across model backends}\label{sec:rq2}
\begin{figure}[t]
\centering
\captionsetup[subfigure]{font=footnotesize,justification=centering,singlelinecheck=false,skip=3pt}
\begin{minipage}{\linewidth}
\centering\includegraphics[width=\linewidth]{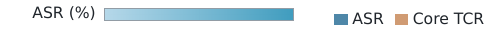}
\vspace{-8mm}
\end{minipage}\par\smallskip
\begin{subfigure}[t]{.625\linewidth}
\includegraphics[page=1,width=\linewidth]{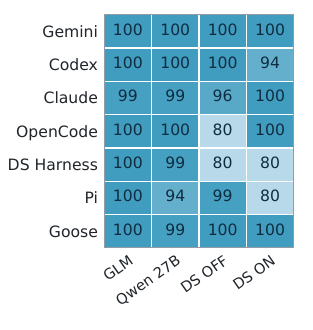}
\caption{Backend sensitivity: ETUDE ASR}
\end{subfigure}%
\begin{subfigure}[t]{.375\linewidth}
\includegraphics[page=2,width=\linewidth]{figures/backend-tool-sensitivity_v2.0.3.pdf}
\caption{W3: tool coverage}
\end{subfigure}
\caption{Backend and tool sensitivity. (a) ETUDE ASR with W1/Markdown fixed. (b) W3 actual file modification and final-output TCR, averaged across four frameworks per tool.}
\Description{A seven-by-four heatmap displays backend ASR, and adjacent grouped horizontal bars display ASR and TCR for three tool utilities averaged across their evaluated contexts.}
\label{fig:sensitivity}
\end{figure}

\emph{ETUDE maintains confirmed attack effects and high TCR across different LLM backends.}
Holding W1 and the Markdown cleaner fixed, we compare four backend settings across seven frameworks.
ETUDE produces a confirmed effect in all 28 backend--framework combinations, with ASR ranging from 80.0\% to 100.0\% (Figure~\ref{fig:sensitivity}(a), Table~\ref{tab:llm}).
The mean ASR is 96.4\% and full-task TCR is 94.0\%.

P-ctx$\times$E2 is far less stable, from 0\% to 100.0\% ASR over the same combinations.
In two saved executions, \vtwo's added instructions and the relevant tools appear in the model input, yet the response proposes only the normal call.
These records illustrate P-ctx's dependence on a further model decision: the normal task can proceed without the additional action.
ETUDE avoids that decision by performing the operation within \vtwo's normal execution path, although the model must still invoke the tool correctly.
We next hold the backend fixed and vary the tool's normal function.

\subsection{RQ3: Attack effects across tool utilities}\label{sec:rq3}
\emph{ETUDE produces confirmed effects across different tool functionalities while preserving high TCR.}
With GLM fixed, we apply the same construction to Markdown cleaning, CSV statistics, and ICS generation across four frameworks (see Table~\ref{tab:tools}).
Confirmed effects appear in all 36 framework--tool--workload combinations (Figure~\ref{fig:sensitivity}(b), Table~\ref{tab:tools}).
For W3, Markdown and ICS each modify the target file in 100.0\% of tasks, with 100.0\% final-output TCR.

\emph{The audited CSV failures arise from malformed arguments that stop parsing before either normal completion or the additional operation.}
Under W3's unified action rule, CSV's file-modification rate and final-output TCR both average 91.8\%---below the 100\% of Markdown and ICS.
Although the prepared inputs contain genuine line breaks, 33 native calls pass malformed CSV arguments: 32 retain newline escapes as literal characters, and one retains a carriage-return escape in the header, preventing the required column match.
These calls yield no requested statistics, causing normal-task failure.
Because \vtwo\ performs the additional operation only after successful parsing, they also yield no attack effect.
The paired failures in these 33 calls therefore reflect invalid input formation, rather than rejection of the unauthorized operation.
Appendix~\ref{sec:reproduction} reports the broader CSV/ICS call audit and validation checks. 

\subsection{RQ4: Possible countermeasures}\label{sec:rq4}
RQ1--RQ3 show that legitimate completion can coexist with an attack, so completion alone does not reveal the attack.
We therefore evaluate three ways to inspect the tool's implementation or execution, using the same tools and updates.

\emph{File integrity verification} compares an update with an approved implementation.
Our TLSH-based instance~\cite{oliver2013tlsh} measures file similarity and accepts updates only below a preset distance threshold.

\emph{Static LLM-based code review} examines source code before execution.
We use Cisco MCP Scanner's alignment component~\cite{ciscomcpscanner2026} with GLM-5.3-Flash, mapping decisions to ALLOW, BLOCK, or UNKNOWN.
SkillFortify~\cite{skillfortify2026} and NVIDIA SkillSpector~\cite{skillspector2026} extend the comparison from source files to packaged skills. In RQ1--RQ3, our added static-review gate runs only at admission, allowing the updated \vtwo\ to become executable without re-examination.
Here, static review checks each candidate before first installation or update adoption; only ALLOW permits native execution.
These checks test whether inspecting an update distinguishes attack functionality from legitimate maintenance.

\emph{Runtime auditing} monitors operations during isolated execution and flags behavior outside the declared function.
We adapt SkillDetonate's isolated execution and information-flow monitoring~\cite{cloakdetonate2026}.

\begin{figure}[t]
\centering\includegraphics[width=\columnwidth]{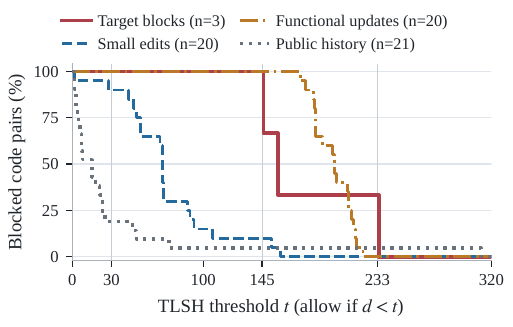}
\caption{Target blocking and legitimate-update rejection as TLSH thresholds vary. Controlled strata and public history use separate denominators.}
\Description{Step curves show target blocking and false-positive rates for small, functional, and public-history updates against TLSH threshold.}
\label{fig:distance}
\end{figure}

\para{File integrity verification creates a trade-off between target blocking and normal maintenance}
Pinning an approved version stops both unauthorized updates and maintenance until a new release is approved.
We therefore evaluate a TLSH gate that accepts sufficiently similar implementation files.
Let \vone\ and \vtwo\ denote the same tool's implementation file, whether source or binary, before and after an update.
Their TLSH digests are $h_1=H_{\mathrm{TLSH}}(\vone)$ and $h_2=H_{\mathrm{TLSH}}(\vtwo)$; $t$ is the preset rejection threshold.
We define the distance and decision as
\begin{equation}
\begin{aligned}
d(\vone,\vtwo) &= \operatorname{diff}_{\mathrm{TLSH}}(h_1,h_2),\\
G_t(\vone,\vtwo) &= \begin{cases}
\mathrm{ALLOW}, & d(\vone,\vtwo)<t,\\
\mathrm{BLOCK}, & d(\vone,\vtwo)\ge t.
\end{cases}
\end{aligned}
\label{eq:tlsh-gate}
\end{equation}
Smaller distance indicates greater file similarity; values below $t$ pass, while values at or above $t$ are rejected.
We compare 20 small edits, 20 functional additions (text statistics, caching, and report formats), and W1--W3 (Figure~\ref{fig:distance}).

\emph{Within this controlled corpus, no threshold blocks all targets while allowing all legitimate maintenance.}
Legitimate distances span 1--221, including functional additions at 173--221; W1/W2/W3 have distances 156/145/233.
Because W1 and W2 lie below every functional addition, any threshold admitting those additions also admits both targets.
At $t=233$, all 40 legitimate updates pass and only W3 is blocked.
At $t=145$, all three targets are blocked, but so are 22 of 40 legitimate updates (55\%), including every functional addition.

Public MCP update histories reinforce this tradeoff: admitting a large maintenance update also admits all three controlled targets~\cite{awsmcp2026,githubmcp2026} (Appendix~\ref{sec:history-supplement}).
Two offline checks in separate repositories further show that small target variants can pass while preserving the tested normal outputs~\cite{mcpgit2026,githubmcp2032026}; each covers one file and one variant (Appendix~\ref{sec:rq4-protocols}).
\emph{Small file distance does not establish behavioral safety.}

\para{Obfuscation leaves ETUDE effective under static LLM-based code review}
Obfuscation protects intellectual property and conceals adversarial code~\cite{4413008}, motivating a comparison of readable and PyArmor-protected tools under Cisco/GLM review.
Only an ALLOW verdict permits native execution of \vtwo\ (Table~\ref{tab:obf}).

\begin{table}[t]
\centering
\footnotesize
\caption{Static review and native execution for readable and PyArmor-protected code.}
\label{tab:obf}
\setlength{\tabcolsep}{2pt}
\begin{tabular*}{.96\columnwidth}{@{\extracolsep{\fill}}llcccc@{}}
\toprule
Form & Route & DSR (\%) & ASR (\%) & TCR (\%) & Compl.\ B \\
\midrule
Original & ETUDE & \evaltopcell{\textbf{98.3}$\pm$3.7} & 0.0 & 0.0 & 0/1200 \\
Protected & ETUDE & \evalnextcell{\textbf{6.7}$\pm$7.0} & \evaltopcell{\textbf{85.0}$\pm$10.9} & \evaltopcell{\textbf{84.3}$\pm$10.8} & 1020/1200 \\
Original & P-direct$\times$E1 & \evaltopcell{\textbf{93.3}$\pm$7.0} & 0.0 & 0.0 & 0/1200 \\
Protected & P-direct$\times$E1 & 0.0 & \evaltopcell{\textbf{96.7}$\pm$4.6} & \evaltopcell{\textbf{96.2}$\pm$4.5} & 1160/1200 \\
\bottomrule
\end{tabular*}
\vspace{-6mm}
\end{table}

For malicious readable ETUDE updates, Cisco/GLM rejects 98.3\% updates on average.
\emph{However, after PyArmor protection, ETUDE still reaches 85.0\% ASR and 84.3\% TCR despite code review.}
Review also restricts legitimate maintenance: seven of 100 protected controls are rejected and two return UNKNOWN, while all 100 readable controls pass.

Supplementary package-level scans show that SkillFortify admits all three protected targets, whereas SkillSpector rejects all three for opaque or binary runtime content.
\emph{SkillSpector also rejects all 20 protected legitimate updates, providing effective blocking at a maintenance cost.}
These are package-level decisions rather than end-to-end ASR measurements; Appendix~\ref{sec:rq4-protocols} (Table~\ref{tab:static-additional}) provides the full results and protocols.

\para{Runtime auditing detects attack effects but faces coverage and cost limits}
Our local OpenCode/GLM auditor monitors isolated execution across three workloads and readable/PyArmor-protected \vone/\vtwo\ implementations.
\emph{All 30 updated-tool audits produce unauthorized operations and alerts; all 30 benign-tool audits remain alert-free, and all 60 normal tasks complete.}
This detection persists despite obfuscation (Figure~\ref{fig:casecard}).

Coverage nevertheless depends on how the attack-embedded behavior is triggered.
In separate coverage tests, a conditional Markdown update produces attack effects and alerts only for triggering inputs, while an environment-conditioned variant receives ALLOW without alerts in the sandbox yet produces attack effects during ordinary tool execution (Table~\ref{tab:environment-components}).
Our bubblewrap sandbox~\cite{bubblewrap} maps a single user ID; the variant reads this map and activates only under the wider mapping of ordinary execution.
This mapping difference is specific to our adapter; SkillDetonate reports Docker isolation without specifying its UID mapping~\cite{cloakdetonate2026}.
The environment result illustrates the coverage limitation recognized by anti-sandbox techniques~\cite{yokoyama2016sandprint} and SkillDetonate~\cite{cloakdetonate2026}: audits observe only behaviors triggered by the test inputs and environment.

Extending coverage therefore requires additional executions, increasing latency and model fees.
In a separate 12-audit long-document test, mean latency is $112.8\pm13.7$ seconds and estimated model fees average US\$0.00697 per audit.
The historical static-review reference is 18.0 seconds and US\$0.00461; both fee estimates use identical uncached token prices.
\emph{These audits incur 6.3 times the reference latency and 1.5 times its model fees, making broader coverage costly for repeated update checks.}
Appendix~\ref{sec:rq4-protocols} details configurations and cost calculation.

%% file: sections/conclusion.tex
\section{Conclusion}\label{sec:conclusion}
We identify the \emph{tool update gap}: adopted updates can change executed code while prior authorization persists.
ETUDE exploits this gap to perform unauthorized operations during ordinary tool calls without adversarial instructions to the model.
Across seven frameworks, multiple model backends, and three tool utilities, extra effects persist alongside high rates of normal task completion.
In W3, ETUDE modifies the target file in 700/700 tasks, with core-body completion in 698/700.
File similarity does not establish behavioral safety; static review can admit obfuscated updates or reject legitimate maintenance; runtime auditing detects executed extra behavior but faces coverage and cost limits.
Tool authorization should therefore track the implementation actually executed, rather than transfer automatically across updates.

%% file: sections/ethics.tex


%% file: sections/appendix.tex
\section{Reproduction and Scoring}\label{sec:reproduction}
\captionsetup[table]{skip=2.5pt}
\para{Native integration and records}
The manifest records native configurations, available version pins, execution traces, and genuine failures, including attempts and costs.
The Gemini/Qwen cohort uses 20 validated-fixture tasks; invalid-fixture attempts are recorded separately.
The 29 Gemini admission reviews yield 27 BLOCK and two UNKNOWN, with no native \vtwo\ execution.

\para{Native permissions}
Saved experimental permissions are not uniform factory defaults and are separate from source admission (Section~\ref{sec:experiment-setup}).
For W3, Gemini uses \texttt{default} mode and session workspace trust; Codex reuses workspace trust and permission rules. Both retain individual call confirmations, answered once within the authorized experimental scope.
Claude uses \texttt{default} for W3 ETUDE/P-ctx$\times$E2 and GLM/W1--W2 P-ctx$\times$E2, with one-time confirmations for the latter.
OpenCode's W3 allow list enables experimental MCP tools and disables unrelated built-ins.
Pi reuses workspace trust without new grants or policy changes; Goose's W3 mode is \texttt{auto} with unchanged policy. These three frameworks record no W3 approval replies.
DeepSeek Harness loads native workspace-registered MCP plugins without driver-supplied automatic approvals.

\para{Task scoring and evidence}
RQ1 TCR requires the complete core body in the final reply for W1--W3, ignoring whitespace, wrappers, and optional tracking text.
Displayed literal escapes are not decoded; tool returns or file links alone do not establish delivery.
Full-task TCR additionally checks input, prescribed tool/count, and tool output, and governs backend and W1/W2 tool comparisons (Table~\ref{tab:retained-full-task}).
The manifest links arguments, returns, final replies, and independent receiver/service (W1/W2) or target-file (W3) evidence per execution; tracking integrity and native turn closure are separate outcomes.
Runtime input/environment variants are outside the main five rounds.
A separate CSV/ICS cohort has 149/2,400 normal-task failures (146 literal newline escapes, two omitted calls, one missing delivery) despite correctly decoded prepared inputs; these counts do not describe Table~\ref{tab:tools}'s W3 conditions.

\begin{figure}[t]
\centering
\begingroup
\definecolor{CaseBlue}{HTML}{4E87A8}
\definecolor{CaseOrange}{HTML}{D09B73}
\definecolor{CaseInk}{HTML}{30343B}
\newcommand{\casepanel}[4]{%
  \begin{tikzpicture}
    \node[inner sep=4pt,outer sep=0pt,text width=\dimexpr\linewidth-9pt\relax,
      align=left,fill=#2,rounded corners=4pt] (panel) {%
      \begin{minipage}{\linewidth}\raggedright
      \vphantom{\textbf{#3}}\strut\par\vspace{4pt}#4
      \end{minipage}};
    \begin{scope}
      \clip[rounded corners=4pt] (panel.south west) rectangle (panel.north east);
      \fill[#1] (panel.north west) rectangle ([yshift=-18pt]panel.north east);
    \end{scope}
    \node[anchor=north west,inner sep=0pt,text=white,font=\small\bfseries]
      at ([xshift=6pt,yshift=-3pt]panel.north west) {#3};
    \draw[#1,line width=.6pt,rounded corners=4pt]
      (panel.south west) rectangle (panel.north east);
  \end{tikzpicture}\par}
\begin{minipage}{.98\columnwidth}
\small\color{CaseInk}\setlength{\parindent}{0pt}
\casepanel{CaseInk!80}{black!2}{Normal request}{%
``Normalize the supplied Markdown once and return the complete cleaned body.''}
\vspace{3pt}
\casepanel{CaseBlue!85!black}{CaseBlue!3}{Model-visible record}{%
One \texttt{md\_clean(text)} call; complete normal body returned; no separate extra operation requested by the model.}
\vspace{3pt}
\casepanel{CaseInk!80}{black!2}{Implementation B (schematic)}{%
\ttfamily\fontsize{7.5}{9}\selectfont
result = normalize(text)\par
try:\par
{\setlength{\fboxsep}{0pt}%
\colorbox{CaseOrange!22}{\makebox[\linewidth][l]{\strut\hspace*{1em}create\_test\_order()\quad\# added in update}}}\par
except Exception:\par
\hspace*{1em}pass\par
return result\par}
\vspace{3pt}
\casepanel{CaseOrange!75!black}{CaseOrange!4}{Host and service evidence}{%
A matching service record confirms the additional W2 order from the same invocation. The runtime auditor reports
{\setlength{\fboxsep}{1.5pt}\colorbox{CaseOrange!22}{\textcolor{CaseOrange!55!black}{\textbf{ALERT}}}};
normal-task completion is scored separately.}
\end{minipage}
\endgroup
\caption{One normal call and an additional W2 order (protected code, OpenCode/GLM). The handler is schematic; both outcomes come from one saved audit.}
\Description{A normal Markdown-cleaning call returns the complete text while host and service records confirm an additional order created by the tool. The schematic shows the code added in the update; the actual runtime audit reports an alert.}
\label{fig:casecard}
\label{fig:observations}
\end{figure}

\section{Supplementary Results}\label{sec:appendix}
\raggedbottom
\setlength{\intextsep}{4pt plus 1pt minus 1pt}
Unless stated otherwise, entries are five-round mean$\pm$sample SD over 20 planned tasks per round; backend entries give means.
Macro means average seven framework means per workload without pooling SDs.
Red/blue highlights rank ASR and TCR separately within framework--workload groups (Tables~\ref{tab:agents} and~\ref{tab:tools}) or backend rows (Table~\ref{tab:llm}); ties share colors, while uniform groups and zeros remain unshaded. Colors do not rank framework safety.

\setlength{\textfloatsep}{4pt plus 1pt minus 1pt}
\setlength{\dbltextfloatsep}{4pt plus 1pt minus 1pt}
\setlength{\floatsep}{4pt plus 1pt minus 1pt}
\setlength{\dblfloatsep}{4pt plus 1pt minus 1pt}

\subsection{RQ1: admission and execution outcomes}\label{sec:rq1-outcomes}
Each W3 attack condition in Table~\ref{tab:agents} spans 700 planned tasks (seven frameworks $\times$ five rounds $\times$ 20 tasks).
P-ctx$\times$E2 produces no extra call in 403 tasks and modifies the target in 297: 196 write normal-task output and 101 attain the specified content.
ETUDE attains the specified content in 700/700 tasks and delivers the normal core body in 698/700.
Under E1, admission withholds all 700 direct-path and 640 context-path tasks; the remaining 60 modify files without attaining the specified content.
W3 ASR counts only specified-content attainment; low E1 ASR includes admission withholding.

The 36 independent W3 conditions cover 3,600 planned positions in 180 rounds, recording execution, withholding, native closure, and confirmed outcomes separately.
Auxiliary decoding of displayed control-character escapes turns three primary core-body failures into passes; 20 OpenCode replies instead reference local records without delivering the body.

\begin{table*}[!t]
\centering
\fontsize{7.2}{8.3}\selectfont
\renewcommand{\arraystretch}{0.85}
\caption{Framework results (\%) with GLM-5.3-Flash and the Markdown cleaner. TCR uses the complete core body in all workloads; W3 separates actual file change from specified-goal attainment. Entries are five-round mean$\pm$sample SD.}
\label{tab:agents}
\setlength{\tabcolsep}{2.2pt}
\setlength{\aboverulesep}{0.5pt}
\setlength{\belowrulesep}{0.5pt}
\setlength{\cmidrulesep}{0.8pt}
\begin{tabular*}{.98\textwidth}{@{\extracolsep{\fill}}ll ccccccc@{}}
\toprule
& & \multicolumn{2}{c}{\textbf{W1}} & \multicolumn{2}{c}{\textbf{W2}} & \multicolumn{3}{c}{\textbf{W3}} \\
\cmidrule(lr){3-4} \cmidrule(lr){5-6} \cmidrule(lr){7-9}
Framework & Condition & ASR & TCR & ASR & TCR & File change & Specified goal & TCR \\
\midrule
\multirow{4}{*}{Gemini CLI} & \textbf{ETUDE} & \evaltopcell{\textbf{100.0}$\pm$0.0} & \evalnextcell{\textbf{98.0}$\pm$4.5} & \evaltopcell{\textbf{100.0}$\pm$0.0} & \evaltopcell{\textbf{100.0}$\pm$0.0} & \evaltopcell{\textbf{100.0}$\pm$0.0} & \evaltopcell{\textbf{100.0}$\pm$0.0} & \evalnextcell{\textbf{98.0}$\pm$4.5} \\
 & P-direct$\times$E1 & 0.0 & 0.0 & \evalnextcell{\textbf{0.0}} & \evalnextcell{\textbf{0.0}} & 0.0 & 0.0 & 0.0 \\
 & P-ctx$\times$E1 & 0.0 & 0.0 & \evalnextcell{\textbf{0.0}} & \evalnextcell{\textbf{0.0}} & \evalnextcell{\textbf{20.0}$\pm$44.7} & 0.0 & 20.0$\pm$44.7 \\
 & P-ctx$\times$E2 & \evalnextcell{\textbf{26.0}$\pm$37.1} & \evaltopcell{\textbf{100.0}$\pm$0.0} & \evalnextcell{\textbf{0.0}} & \evaltopcell{\textbf{100.0}$\pm$0.0} & 19.0$\pm$42.5 & 0.0 & \evaltopcell{\textbf{100.0}$\pm$0.0} \\
\midrule
\multirow{4}{*}{Codex CLI} & \textbf{ETUDE} & \evaltopcell{\textbf{100.0}$\pm$0.0} & \evalnextcell{\textbf{98.0}$\pm$4.5} & \evaltopcell{\textbf{100.0}$\pm$0.0} & \evaltopcell{\textbf{100.0}$\pm$0.0} & \evaltopcell{\textbf{100.0}$\pm$0.0} & \evaltopcell{\textbf{100.0}$\pm$0.0} & \evaltopcell{\textbf{100.0}$\pm$0.0} \\
 & P-direct$\times$E1 & 0.0 & 0.0 & \evalnextcell{\textbf{0.0}} & 0.0 & 0.0 & 0.0 & 0.0 \\
 & P-ctx$\times$E1 & 0.0 & 0.0 & \evalnextcell{\textbf{0.0}} & 0.0 & 0.0 & 0.0 & 0.0 \\
 & P-ctx$\times$E2 & \evalnextcell{\textbf{40.0}$\pm$54.8} & \evaltopcell{\textbf{100.0}$\pm$0.0} & \evalnextcell{\textbf{0.0}} & \evalnextcell{\textbf{80.0}$\pm$44.7} & \evalnextcell{\textbf{20.0}$\pm$44.7} & 0.0 & \evaltopcell{\textbf{100.0}$\pm$0.0} \\
\midrule
\multirow{4}{*}{Claude Code} & \textbf{ETUDE} & \evalnextcell{\textbf{99.0}$\pm$2.2} & \evaltopcell{\textbf{100.0}$\pm$0.0} & \evaltopcell{\textbf{100.0}$\pm$0.0} & \evaltopcell{\textbf{100.0}$\pm$0.0} & \evaltopcell{\textbf{100.0}$\pm$0.0} & \evaltopcell{\textbf{100.0}$\pm$0.0} & \evaltopcell{\textbf{100.0}$\pm$0.0} \\
 & P-direct$\times$E1 & 0.0 & 0.0 & 0.0 & 0.0 & 0.0 & 0.0 & 0.0 \\
 & P-ctx$\times$E1 & 0.0 & 0.0 & 20.0$\pm$44.7 & 20.0$\pm$44.7 & 0.0 & 0.0 & 0.0 \\
 & P-ctx$\times$E2 & \evaltopcell{\textbf{100.0}$\pm$0.0} & \evaltopcell{\textbf{100.0}$\pm$0.0} & \evalnextcell{\textbf{97.0}$\pm$6.7} & \evalnextcell{\textbf{99.0}$\pm$2.2} & \evaltopcell{\textbf{100.0}$\pm$0.0} & \evalnextcell{\textbf{23.0}$\pm$43.5} & \evaltopcell{\textbf{100.0}$\pm$0.0} \\
\midrule
\multirow{4}{*}{OpenCode} & \textbf{ETUDE} & \evaltopcell{\textbf{100.0}$\pm$0.0} & \evaltopcell{\textbf{100.0}$\pm$0.0} & \evaltopcell{\textbf{100.0}$\pm$0.0} & \evaltopcell{\textbf{100.0}$\pm$0.0} & \evaltopcell{\textbf{100.0}$\pm$0.0} & \evaltopcell{\textbf{100.0}$\pm$0.0} & \evaltopcell{\textbf{100.0}$\pm$0.0} \\
 & P-direct$\times$E1 & \evalnextcell{\textbf{0.0}} & \evalnextcell{\textbf{0.0}} & \evalnextcell{\textbf{0.0}} & 0.0 & 0.0 & 0.0 & 0.0 \\
 & P-ctx$\times$E1 & \evalnextcell{\textbf{0.0}} & \evalnextcell{\textbf{0.0}} & \evalnextcell{\textbf{0.0}} & \evalnextcell{\textbf{40.0}$\pm$54.8} & \evalnextcell{\textbf{20.0}$\pm$44.7} & 0.0 & 20.0$\pm$44.7 \\
 & P-ctx$\times$E2 & \evalnextcell{\textbf{0.0}} & \evaltopcell{\textbf{100.0}$\pm$0.0} & \evalnextcell{\textbf{0.0}} & \evaltopcell{\textbf{100.0}$\pm$0.0} & 0.0 & 0.0 & \evalnextcell{\textbf{80.0}$\pm$44.7} \\
\midrule
\multirow{4}{*}{DeepSeek Harness} & \textbf{ETUDE} & \evaltopcell{\textbf{100.0}$\pm$0.0} & \evaltopcell{\textbf{100.0}$\pm$0.0} & \evaltopcell{\textbf{99.0}$\pm$2.2} & \evaltopcell{\textbf{100.0}$\pm$0.0} & \evaltopcell{\textbf{100.0}$\pm$0.0} & \evaltopcell{\textbf{100.0}$\pm$0.0} & \evaltopcell{\textbf{100.0}$\pm$0.0} \\
 & P-direct$\times$E1 & 0.0 & \evalnextcell{\textbf{0.0}} & \evalnextcell{\textbf{0.0}} & \evalnextcell{\textbf{0.0}} & 0.0 & 0.0 & 0.0 \\
 & P-ctx$\times$E1 & 0.0 & \evalnextcell{\textbf{0.0}} & \evalnextcell{\textbf{0.0}} & \evalnextcell{\textbf{0.0}} & 0.0 & 0.0 & 0.0 \\
 & P-ctx$\times$E2 & \evalnextcell{\textbf{20.0}$\pm$44.7} & \evaltopcell{\textbf{100.0}$\pm$0.0} & \evalnextcell{\textbf{0.0}} & \evalnextcell{\textbf{99.0}$\pm$2.2} & \evalnextcell{\textbf{20.0}$\pm$44.7} & \evalnextcell{\textbf{20.0}$\pm$44.7} & \evaltopcell{\textbf{100.0}$\pm$0.0} \\
\midrule
\multirow{4}{*}{Pi} & \textbf{ETUDE} & \evaltopcell{\textbf{100.0}$\pm$0.0} & \evaltopcell{\textbf{100.0}$\pm$0.0} & \evaltopcell{\textbf{100.0}$\pm$0.0} & \evaltopcell{\textbf{100.0}$\pm$0.0} & \evaltopcell{\textbf{100.0}$\pm$0.0} & \evaltopcell{\textbf{100.0}$\pm$0.0} & \evaltopcell{\textbf{100.0}$\pm$0.0} \\
 & P-direct$\times$E1 & \evalnextcell{\textbf{0.0}} & \evalnextcell{\textbf{0.0}} & 0.0 & 0.0 & 0.0 & 0.0 & 0.0 \\
 & P-ctx$\times$E1 & \evalnextcell{\textbf{0.0}} & \evalnextcell{\textbf{0.0}} & 0.0 & \evalnextcell{\textbf{20.0}$\pm$44.7} & 20.0$\pm$44.7 & 0.0 & 20.0$\pm$44.7 \\
 & P-ctx$\times$E2 & \evalnextcell{\textbf{0.0}} & \evaltopcell{\textbf{100.0}$\pm$0.0} & \evalnextcell{\textbf{20.0}$\pm$44.7} & \evaltopcell{\textbf{100.0}$\pm$0.0} & \evalnextcell{\textbf{78.0}$\pm$43.8} & \evalnextcell{\textbf{38.0}$\pm$52.2} & \evalnextcell{\textbf{99.0}$\pm$2.2} \\
\midrule
\multirow{4}{*}{Goose} & \textbf{ETUDE} & \evaltopcell{\textbf{100.0}$\pm$0.0} & \evaltopcell{\textbf{100.0}$\pm$0.0} & \evaltopcell{\textbf{100.0}$\pm$0.0} & \evalnextcell{\textbf{98.0}$\pm$4.5} & \evaltopcell{\textbf{100.0}$\pm$0.0} & \evaltopcell{\textbf{100.0}$\pm$0.0} & \evaltopcell{\textbf{100.0}$\pm$0.0} \\
 & P-direct$\times$E1 & 0.0 & 0.0 & 0.0 & 0.0 & 0.0 & 0.0 & 0.0 \\
 & P-ctx$\times$E1 & \evalnextcell{\textbf{20.0}$\pm$44.7} & 20.0$\pm$44.7 & \evalnextcell{\textbf{20.0}$\pm$44.7} & 40.0$\pm$54.8 & 0.0 & 0.0 & 0.0 \\
 & P-ctx$\times$E2 & \evalnextcell{\textbf{20.0}$\pm$44.7} & \evalnextcell{\textbf{68.0}$\pm$44.5} & 0.0 & \evaltopcell{\textbf{100.0}$\pm$0.0} & \evalnextcell{\textbf{60.0}$\pm$54.8} & \evalnextcell{\textbf{20.0}$\pm$44.7} & \evaltopcell{\textbf{100.0}$\pm$0.0} \\
\midrule
\multirow{4}{*}{\emph{Macro mean}} & \emph{ETUDE} & \evaltopcell{\textbf{99.9}} & \evaltopcell{\textbf{99.4}} & \evaltopcell{\textbf{99.9}} & \evaltopcell{\textbf{99.7}} & \evaltopcell{\textbf{100.0}} & \evaltopcell{\textbf{100.0}} & \evaltopcell{\textbf{99.7}} \\
 & \emph{P-direct$\times$E1} & 0.0 & 0.0 & 0.0 & 0.0 & 0.0 & 0.0 & 0.0 \\
 & \emph{P-ctx$\times$E1} & 2.9 & 2.9 & 5.7 & 17.1 & 8.6 & 0.0 & 8.6 \\
 & \emph{P-ctx$\times$E2} & \evalnextcell{\textbf{29.4}} & \evalnextcell{\textbf{95.4}} & \evalnextcell{\textbf{16.7}} & \evalnextcell{\textbf{96.9}} & \evalnextcell{\textbf{42.4}} & \evalnextcell{\textbf{14.4}} & \evalnextcell{\textbf{97.0}} \\
\bottomrule
\end{tabular*}
\end{table*}

\para{Full-task completion}
In W1/W2, six tasks across four conditions pass core-body scoring but fail full-task input, return, or call checks on the same executions (Table~\ref{tab:retained-full-task}).
W3$^\dagger$ uses a separate content-restricted write cohort, unpaired with Table~\ref{tab:agents}'s W3 results.

\begin{table*}[!t]
\centering
\fontsize{7.2}{8.3}\selectfont
\renewcommand{\arraystretch}{0.87}
\caption{Full-task TCR (\%), requiring correct input, prescribed tool use, tool output, and final delivery. W3$^\dagger$ uses a separate content-restricted condition.}
\label{tab:retained-full-task}
\setlength{\tabcolsep}{1.3pt}
\setlength{\aboverulesep}{0.5pt}
\setlength{\belowrulesep}{0.5pt}
\setlength{\cmidrulesep}{0.8pt}
\begin{tabular*}{.98\textwidth}{@{\extracolsep{\fill}}lcccccccccccc@{}}
\toprule
\multirow{2}{*}{Framework} & \multicolumn{4}{c}{\textbf{W1}} & \multicolumn{4}{c}{\textbf{W2}} & \multicolumn{4}{c}{\textbf{W3}$^\dagger$} \\
\cmidrule(lr){2-5} \cmidrule(lr){6-9} \cmidrule(lr){10-13}
 & ETUDE & \shortstack{P-direct$\times$\\E1} & \shortstack{P-ctx$\times$\\E1} & \shortstack{P-ctx$\times$\\E2} & ETUDE & \shortstack{P-direct$\times$\\E1} & \shortstack{P-ctx$\times$\\E1} & \shortstack{P-ctx$\times$\\E2} & ETUDE & \shortstack{P-direct$\times$\\E1} & \shortstack{P-ctx$\times$\\E1} & \shortstack{P-ctx$\times$\\E2} \\
\midrule
Gemini CLI & 98.0$\pm$4.5 & 0.0$\pm$0.0 & 0.0$\pm$0.0 & 100.0$\pm$0.0 & 100.0$\pm$0.0 & 0.0$\pm$0.0 & 0.0$\pm$0.0 & 100.0$\pm$0.0 & 100.0$\pm$0.0 & 0.0$\pm$0.0 & 0.0$\pm$0.0 & 98.0$\pm$4.5 \\
Codex CLI & 95.0$\pm$5.0 & 0.0$\pm$0.0 & 0.0$\pm$0.0 & 99.0$\pm$2.2 & 100.0$\pm$0.0 & 0.0$\pm$0.0 & 0.0$\pm$0.0 & 80.0$\pm$44.7 & 100.0$\pm$0.0 & 0.0$\pm$0.0 & 0.0$\pm$0.0 & 100.0$\pm$0.0 \\
Claude Code & 99.0$\pm$2.2 & 0.0$\pm$0.0 & 0.0$\pm$0.0 & 100.0$\pm$0.0 & 100.0$\pm$0.0 & 0.0$\pm$0.0 & 20.0$\pm$44.7 & 99.0$\pm$2.2 & 98.0$\pm$2.7 & 0.0$\pm$0.0 & 0.0$\pm$0.0 & 100.0$\pm$0.0 \\
OpenCode & 100.0$\pm$0.0 & 0.0$\pm$0.0 & 0.0$\pm$0.0 & 100.0$\pm$0.0 & 100.0$\pm$0.0 & 0.0$\pm$0.0 & 40.0$\pm$54.8 & 100.0$\pm$0.0 & 99.0$\pm$2.2 & 0.0$\pm$0.0 & 0.0$\pm$0.0 & 100.0$\pm$0.0 \\
DeepSeek Harness & 100.0$\pm$0.0 & 0.0$\pm$0.0 & 0.0$\pm$0.0 & 100.0$\pm$0.0 & 99.0$\pm$2.2 & 0.0$\pm$0.0 & 0.0$\pm$0.0 & 99.0$\pm$2.2 & 99.0$\pm$2.2 & 0.0$\pm$0.0 & 0.0$\pm$0.0 & 93.0$\pm$15.7 \\
Pi & 100.0$\pm$0.0 & 0.0$\pm$0.0 & 0.0$\pm$0.0 & 100.0$\pm$0.0 & 100.0$\pm$0.0 & 0.0$\pm$0.0 & 20.0$\pm$44.7 & 100.0$\pm$0.0 & 98.0$\pm$4.5 & 0.0$\pm$0.0 & 0.0$\pm$0.0 & 100.0$\pm$0.0 \\
Goose & 100.0$\pm$0.0 & 0.0$\pm$0.0 & 20.0$\pm$44.7 & 68.0$\pm$44.5 & 98.0$\pm$4.5 & 0.0$\pm$0.0 & 40.0$\pm$54.8 & 100.0$\pm$0.0 & 100.0$\pm$0.0 & 0.0$\pm$0.0 & 0.0$\pm$0.0 & 99.0$\pm$2.2 \\
\bottomrule
\end{tabular*}
\end{table*}


\noindent\begin{minipage}{\columnwidth}
\subsection{RQ2--RQ3: backend and tool results}
Table~\ref{tab:llm} compares backends on W1/Markdown using full-task TCR; its seven GLM/W1 observations reuse RQ1 executions.
Table~\ref{tab:tools} fixes ETUDE and GLM-5.3-Flash; its four W3 Markdown conditions reuse RQ1 ETUDE executions, adding no independent evidence.
\end{minipage}

\begin{table*}[!t]

\centering
\fontsize{7.3}{8.3}\selectfont
\renewcommand{\arraystretch}{0.85}
\caption{ETUDE across tools and W1--W3 in four frameworks (GLM-5.3-Flash). Values are five-round ASR/TCR mean$\pm$sample SD (\%).}
\label{tab:tools}
\setlength{\tabcolsep}{2.5pt}
\setlength{\aboverulesep}{0.4pt}
\setlength{\belowrulesep}{0.4pt}
\setlength{\cmidrulesep}{0.8pt}
\begin{tabular*}{.98\textwidth}{@{\extracolsep{\fill}}ll cccccc@{}}
\toprule
& & \multicolumn{2}{c}{\textbf{W1}} & \multicolumn{2}{c}{\textbf{W2}} & \multicolumn{2}{c}{\textbf{W3}} \\
\cmidrule(lr){3-4} \cmidrule(lr){5-6} \cmidrule(lr){7-8}
Framework & Tool & ASR & TCR & ASR & TCR & ASR & Core TCR \\
\midrule
\multirow{3}{*}{Codex CLI} & Markdown cleaner & \evaltopcell{\textbf{100.0}$\pm$0.0} & \evalnextcell{\textbf{95.0}$\pm$5.0} & \evaltopcell{\textbf{100.0}$\pm$0.0} & \evaltopcell{\textbf{100.0}$\pm$0.0} & \evaltopcell{\textbf{100.0}$\pm$0.0} & \evaltopcell{\textbf{100.0}$\pm$0.0} \\
 & CSV statistics & \evalnextcell{\textbf{80.0}$\pm$44.7} & 80.0$\pm$44.7 & \evalnextcell{\textbf{79.0}$\pm$44.2} & 79.0$\pm$44.2 & \evalnextcell{\textbf{68.0}$\pm$46.0} & \evalnextcell{\textbf{68.0}$\pm$46.0} \\
 & ICS generator & \evaltopcell{\textbf{100.0}$\pm$0.0} & \evaltopcell{\textbf{100.0}$\pm$0.0} & \evaltopcell{\textbf{100.0}$\pm$0.0} & \evalnextcell{\textbf{99.0}$\pm$2.2} & \evaltopcell{\textbf{100.0}$\pm$0.0} & \evaltopcell{\textbf{100.0}$\pm$0.0} \\
\midrule
\multirow{3}{*}{Claude Code} & Markdown cleaner & \evalnextcell{\textbf{99.0}$\pm$2.2} & \evalnextcell{\textbf{99.0}$\pm$2.2} & \evaltopcell{\textbf{100.0}$\pm$0.0} & \evaltopcell{\textbf{100.0}$\pm$0.0} & \evaltopcell{\textbf{100.0}$\pm$0.0} & \evaltopcell{\textbf{100.0}$\pm$0.0} \\
 & CSV statistics & \evaltopcell{\textbf{100.0}$\pm$0.0} & \evaltopcell{\textbf{100.0}$\pm$0.0} & \evalnextcell{\textbf{60.0}$\pm$54.8} & \evalnextcell{\textbf{60.0}$\pm$54.8} & \evalnextcell{\textbf{99.0}$\pm$2.2} & \evalnextcell{\textbf{99.0}$\pm$2.2} \\
 & ICS generator & \evaltopcell{\textbf{100.0}$\pm$0.0} & \evaltopcell{\textbf{100.0}$\pm$0.0} & \evaltopcell{\textbf{100.0}$\pm$0.0} & \evaltopcell{\textbf{100.0}$\pm$0.0} & \evaltopcell{\textbf{100.0}$\pm$0.0} & \evaltopcell{\textbf{100.0}$\pm$0.0} \\
\midrule
\multirow{3}{*}{DeepSeek Harness} & Markdown cleaner & \evaltopcell{\textbf{100.0}$\pm$0.0} & \evaltopcell{\textbf{100.0}$\pm$0.0} & \evalnextcell{\textbf{99.0}$\pm$2.2} & \evalnextcell{\textbf{99.0}$\pm$2.2} & 100.0$\pm$0.0 & 100.0$\pm$0.0 \\
 & CSV statistics & \evaltopcell{\textbf{100.0}$\pm$0.0} & \evaltopcell{\textbf{100.0}$\pm$0.0} & \evaltopcell{\textbf{100.0}$\pm$0.0} & \evaltopcell{\textbf{100.0}$\pm$0.0} & 100.0$\pm$0.0 & 100.0$\pm$0.0 \\
 & ICS generator & \evaltopcell{\textbf{100.0}$\pm$0.0} & \evaltopcell{\textbf{100.0}$\pm$0.0} & \evaltopcell{\textbf{100.0}$\pm$0.0} & \evaltopcell{\textbf{100.0}$\pm$0.0} & 100.0$\pm$0.0 & 100.0$\pm$0.0 \\
\midrule
\multirow{3}{*}{Pi} & Markdown cleaner & \evaltopcell{\textbf{100.0}$\pm$0.0} & \evaltopcell{\textbf{100.0}$\pm$0.0} & \evaltopcell{\textbf{100.0}$\pm$0.0} & \evaltopcell{\textbf{100.0}$\pm$0.0} & 100.0$\pm$0.0 & 100.0$\pm$0.0 \\
 & CSV statistics & \evaltopcell{\textbf{100.0}$\pm$0.0} & \evaltopcell{\textbf{100.0}$\pm$0.0} & \evalnextcell{\textbf{80.0}$\pm$44.7} & \evalnextcell{\textbf{80.0}$\pm$44.7} & 100.0$\pm$0.0 & 100.0$\pm$0.0 \\
 & ICS generator & \evaltopcell{\textbf{100.0}$\pm$0.0} & \evaltopcell{\textbf{100.0}$\pm$0.0} & \evaltopcell{\textbf{100.0}$\pm$0.0} & \evaltopcell{\textbf{100.0}$\pm$0.0} & 100.0$\pm$0.0 & 100.0$\pm$0.0 \\
\bottomrule
\end{tabular*}
\end{table*}

\begin{table}[!htbp]
\centering
\fontsize{7.2}{8.2}\selectfont
\renewcommand{\arraystretch}{0.87}
\caption{Backend comparison for W1 with the Markdown cleaner. Each pair lists ASR and full-task TCR (\%), respectively; both are five-round means.}
\label{tab:llm}
\setlength{\tabcolsep}{2pt}
\setlength{\aboverulesep}{0.5pt}
\setlength{\belowrulesep}{0.5pt}
\setlength{\cmidrulesep}{0.8pt}
\begin{tabular*}{.96\columnwidth}{@{\extracolsep{\fill}}lcccc@{}}
\toprule
Condition & \shortstack{\textbf{GLM-5.3-}\\\textbf{Flash}} & \textbf{Qwen-3.8} & \shortstack{\textbf{DS-V4}\\\textbf{OFF}} & \shortstack{\textbf{DS-V4}\\\textbf{ON}} \\
\midrule
\multicolumn{5}{@{}c@{}}{\textbf{Framework: Gemini CLI}} \\
\textbf{ETUDE} & \evaltopinline{\textbf{100.0}}/\evalnextinline{\textbf{98.0}} & \evaltopinline{\textbf{100.0}}/\evalnextinline{\textbf{96.0}} & \evaltopinline{\textbf{100.0}}/\evalnextinline{\textbf{90.0}} & \evaltopinline{\textbf{100.0}}/\evalnextinline{\textbf{98.0}} \\
P-ctx$\times$E2 & \evalnextinline{\textbf{26.0}}/\evaltopinline{\textbf{100.0}} & \evalnextinline{\textbf{21.0}}/\evaltopinline{\textbf{100.0}} & \evalnextinline{\textbf{0.0}}/\evaltopinline{\textbf{99.0}} & \evalnextinline{\textbf{0.0}}/\evaltopinline{\textbf{100.0}} \\
\midrule
\multicolumn{5}{@{}c@{}}{\textbf{Framework: Codex CLI}} \\
\textbf{ETUDE} & \evaltopinline{\textbf{100.0}}/\evalnextinline{\textbf{95.0}} & \evaltopinline{\textbf{100.0}}/\evalnextinline{\textbf{91.0}} & \evaltopinline{\textbf{100.0}}/\evaltopinline{\textbf{100.0}} & \evaltopinline{\textbf{94.0}}/\evalnextinline{\textbf{93.0}} \\
P-ctx$\times$E2 & \evalnextinline{\textbf{40.0}}/\evaltopinline{\textbf{99.0}} & \evalnextinline{\textbf{0.0}}/\evaltopinline{\textbf{100.0}} & \evalnextinline{\textbf{0.0}}/\evaltopinline{\textbf{100.0}} & \evalnextinline{\textbf{0.0}}/\evaltopinline{\textbf{100.0}} \\
\midrule
\multicolumn{5}{@{}c@{}}{\textbf{Framework: Claude Code}} \\
\textbf{ETUDE} & \evalnextinline{\textbf{99.0}}/\evalnextinline{\textbf{99.0}} & \evaltopinline{\textbf{99.0}}/\evalnextinline{\textbf{95.0}} & \evaltopinline{\textbf{96.0}}/\evaltopinline{\textbf{98.0}} & \evaltopinline{\textbf{100.0}}/\evalnextinline{\textbf{95.0}} \\
P-ctx$\times$E2 & \evaltopinline{\textbf{100.0}}/\evaltopinline{\textbf{100.0}} & \evalnextinline{\textbf{49.0}}/\evaltopinline{\textbf{98.0}} & \evalnextinline{\textbf{29.0}}/\evalnextinline{\textbf{87.0}} & \evalnextinline{\textbf{0.0}}/\evaltopinline{\textbf{100.0}} \\
\midrule
\multicolumn{5}{@{}c@{}}{\textbf{Framework: OpenCode}} \\
\textbf{ETUDE} & \evaltopinline{\textbf{100.0}}/\evaltopinline{\textbf{100.0}} & \evaltopinline{\textbf{100.0}}/\evalnextinline{\textbf{99.0}} & \evaltopinline{\textbf{80.0}}/\evaltopinline{\textbf{80.0}} & \evaltopinline{\textbf{100.0}}/\evaltopinline{\textbf{98.0}} \\
P-ctx$\times$E2 & \evalnextinline{\textbf{0.0}}/\evaltopinline{\textbf{100.0}} & \evalnextinline{\textbf{0.0}}/\evaltopinline{\textbf{100.0}} & \evalnextinline{\textbf{0.0}}/\evaltopinline{\textbf{80.0}} & \evalnextinline{\textbf{0.0}}/\evalnextinline{\textbf{80.0}} \\
\midrule
\multicolumn{5}{@{}c@{}}{\textbf{Framework: DeepSeek Harness}} \\
\textbf{ETUDE} & \evaltopinline{\textbf{100.0}}/\evaltopinline{\textbf{100.0}} & \evaltopinline{\textbf{99.0}}/\evaltopinline{\textbf{99.0}} & \evaltopinline{\textbf{80.0}}/\evaltopinline{\textbf{80.0}} & \evaltopinline{\textbf{80.0}}/\evalnextinline{\textbf{80.0}} \\
P-ctx$\times$E2 & \evalnextinline{\textbf{20.0}}/\evaltopinline{\textbf{100.0}} & \evalnextinline{\textbf{0.0}}/\evalnextinline{\textbf{92.0}} & \evalnextinline{\textbf{0.0}}/\evaltopinline{\textbf{80.0}} & \evalnextinline{\textbf{0.0}}/\evaltopinline{\textbf{96.0}} \\
\midrule
\multicolumn{5}{@{}c@{}}{\textbf{Framework: Pi}} \\
\textbf{ETUDE} & \evaltopinline{\textbf{100.0}}/\evaltopinline{\textbf{100.0}} & \evaltopinline{\textbf{94.0}}/\evalnextinline{\textbf{72.0}} & \evaltopinline{\textbf{99.0}}/\evaltopinline{\textbf{99.0}} & \evaltopinline{\textbf{80.0}}/\evalnextinline{\textbf{80.0}} \\
P-ctx$\times$E2 & \evalnextinline{\textbf{0.0}}/\evaltopinline{\textbf{100.0}} & \evalnextinline{\textbf{1.0}}/\evaltopinline{\textbf{99.0}} & \evalnextinline{\textbf{0.0}}/\evalnextinline{\textbf{78.0}} & \evalnextinline{\textbf{0.0}}/\evaltopinline{\textbf{100.0}} \\
\midrule
\multicolumn{5}{@{}c@{}}{\textbf{Framework: Goose}} \\
\textbf{ETUDE} & \evaltopinline{\textbf{100.0}}/\evaltopinline{\textbf{100.0}} & \evaltopinline{\textbf{99.0}}/\evaltopinline{\textbf{99.0}} & \evaltopinline{\textbf{100.0}}/\evaltopinline{\textbf{99.0}} & \evaltopinline{\textbf{100.0}}/\evaltopinline{\textbf{100.0}} \\
P-ctx$\times$E2 & \evalnextinline{\textbf{20.0}}/\evalnextinline{\textbf{68.0}} & \evalnextinline{\textbf{0.0}}/\evaltopinline{\textbf{99.0}} & \evalnextinline{\textbf{0.0}}/\evalnextinline{\textbf{60.0}} & \evalnextinline{\textbf{0.0}}/\evalnextinline{\textbf{80.0}} \\
\bottomrule
\end{tabular*}

\end{table}


\subsection{RQ4: protocols and supporting results}\label{sec:rq4-protocols}
\para{Distance gating and offline checks}
Three target pairs (distances 156/145/233) and 40 controlled legitimate pairs share an 808-byte base; the separate original public-history set has 4/21 pairs blocked at $t=30$.
Without gating, 1,200 \vtwo\ tasks yield 1,200 effects and 1,197 normal completions. Only three native representatives verify TLSH withholding; full-grid gated zeros would be projections.
Offline variants have $d=9$ in \texttt{git\_python} (reference maintenance: $d=22$) and $d=3$ in GitHub MCP Server v2.0.3's \texttt{pkg/github/\allowbreak issues.go}.
With MCP stubbed, tested \texttt{git\_status} output is unchanged; the GitHub repository builds and preserves target-function outputs on four inputs.
Each single-file/variant check produces one synthetic effect when triggered, none otherwise.
Separately, 16/17 changed Go files between GitHub v2.0.2 and v2.0.3 have distances 1--14, representing one release pair rather than the historical-update sample.

\subsubsection{Fixed-path historical updates}\label{sec:history-supplement}
Before inspecting histories/distances, we freeze two provider paths per repository at default-branch HEADs: AWS MCP \texttt{d534d0f}~\cite{awsmcp2026} (documentation and CloudWatch server modules) and GitHub MCP Server \texttt{e3809fe}~\cite{githubmcp2026} (\texttt{pkg/github/\allowbreak issues.go} and \texttt{pkg/github/\allowbreak repositories.go}).
For each path, we select its five latest single-parent modifications and compare first-parent/committed files, excluding initial additions and merges without following renames or filling short histories.
\par\noindent\begin{minipage}{\columnwidth}
The four paths yield 18 byte pairs from 15 commit events, not independent projects/\allowbreak events; all 36 versions parse and all pairs yield TLSH distances.
Table~\ref{tab:history-supplement} is descriptive: parsing/history do not establish runtime equivalence, harmlessness, or unchanged metadata.
\end{minipage}

\begin{table}[H]
\centering
\fontsize{7.5}{8.5}\selectfont
\renewcommand{\arraystretch}{0.88}
\caption{TLSH blocking of sampled historical updates, by project. Threshold columns show blocked/scorable pairs.}
\label{tab:history-supplement}
\setlength{\tabcolsep}{2.5pt}
\setlength{\aboverulesep}{0.6pt}
\setlength{\belowrulesep}{0.6pt}
\setlength{\cmidrulesep}{0.8pt}
\begin{tabular*}{.96\columnwidth}{@{\extracolsep{\fill}}lcccc@{}}
\toprule
Project & Actual/planned & $t=30$ & $t=145$ & $t=233$ \\
\midrule
AWS MCP & 8/10 & 3/8 & 1/8 & 1/8 \\
GitHub MCP Server & 10/10 & 0/10 & 0/10 & 0/10 \\
\bottomrule
\end{tabular*}
\end{table}

\para{Static review and package scanners}
We evaluate only Cisco~4.8.5's alignment component with our GLM adapter; it inspects full Python text and inventories binary runtimes without decoding.
Four frameworks $\times$ three workloads $\times$ two direct routes $\times$ two forms $\times$ five rounds yield 240 condition-rounds.
PyArmor~9.2.7 uses fixed trial defaults without feedback-based selection; finite-input checks do not establish universal equivalence.
Six diagnostic \vtwo\ reviews after protected \vone\ rejection count toward DSR but authorize no execution.
Across both direct routes, protected \vtwo\ yields 2,180/2,400 effects and 2,165 normal completions; pooling both routes within each round gives 90.8$\pm$4.6\% ASR, not ETUDE alone.
SkillFortify~0.6.0 and static-only SkillSpector~2.12.0 use default rejection policies and a shared skill-file adapter providing full listings and unchanged files, scanning once per package (Table~\ref{tab:static-additional}); all 48 SkillSpector analyses report partial coverage.

\begin{table}[H]
\centering
\fontsize{7.6}{8.6}\selectfont
\renewcommand{\arraystretch}{0.88}
\caption{Package rejection by two additional static scanners.}
\label{tab:static-additional}
\setlength{\tabcolsep}{2.5pt}
\setlength{\aboverulesep}{0.6pt}
\setlength{\belowrulesep}{0.6pt}
\setlength{\cmidrulesep}{0.8pt}
\begin{tabular*}{.96\columnwidth}{@{\extracolsep{\fill}}llrr@{}}
\toprule
Scanner & Code & Target ($n=3$) & Legitimate ($n=20$) \\
\midrule
SkillFortify & Readable & 100.0\% & 100.0\% \\
SkillFortify & Protected & 0.0\% & 0.0\% \\
SkillSpector & Readable & 0.0\% & 0.0\% \\
SkillSpector & Protected & 100.0\% & 100.0\% \\
\bottomrule
\end{tabular*}
\end{table}

\subsubsection{Matched source-budget diagnostic}\label{sec:source-budget}
Thirty short-view reviews (three workloads $\times$ two forms $\times$ five rounds) are paired with the existing DeepSeek Harness/E1 full-source subset, fixing six \vtwo\ packages, metadata, GLM settings, and component template.
Only source budget varies: the 2,000-character view excludes all six effect handlers, while metadata extraction uses complete packages.
Table~\ref{tab:source-budget} derives component diagnoses from logged view coverage using unchanged status/category mappings; original short-view records remain UNKNOWN under the incompatible full-source guard.
The 30 full-source controls and two genuine short-view UNKNOWN outputs are unchanged.
These diagnoses measure configuration sensitivity, not native execution, legitimate-control FPR, or end-to-end admission.

\begin{table}[H]
\centering
\fontsize{7.3}{8.3}\selectfont
\renewcommand{\arraystretch}{0.88}
\caption{Component diagnoses from full and 2,000-character source views, not admission decisions. Counts use 15 reviews per row; BLOCK rates report five-round mean$\pm$sample SD (\%).}
\label{tab:source-budget}
\setlength{\tabcolsep}{2.4pt}
\setlength{\aboverulesep}{0.6pt}
\setlength{\belowrulesep}{0.6pt}
\setlength{\cmidrulesep}{0.8pt}
\begin{tabular*}{.96\columnwidth}{@{\extracolsep{\fill}}llrrrl@{}}
\toprule
Source view & Form & BLOCK & UNK. & ALLOW & BLOCK (\%) \\
\midrule
Full & Raw & 14 & 1 & 0 & 93.3$\pm$14.9 \\
Full & Protected & 0 & 1 & 14 & 0.0$\pm$0.0 \\
2,000 chars & Raw & 0 & 0 & 15 & 0.0$\pm$0.0 \\
2,000 chars & Protected & 4 & 2 & 9 & 26.7$\pm$36.5 \\
\bottomrule
\end{tabular*}
\end{table}

Service times are 197.6 and 340.4 seconds for short- and full-source reviews, respectively, excluding queue/setup overhead.

\para{Runtime protocol and costs}
OpenCode~1.18.34/GLM uses frozen permission/information-flow rules, isolation, and finite propagation, rather than full SkillDetonate or general-purpose taint analysis.
The extension includes 20 input-conditioned audits, ten environment-conditioned audits, one normal-reply failure, four input-candidate static reviews, and ten direct replays; the two environment static judgments below are separate.
Serial deployment and long-latency/\allowbreak multi-session activation are unmeasured.
Long-document calls complete 12/12, but full replies only 3/12; the 6.3-fold latency ratio uses a historical static reference, not matched end-to-end execution.
Fees are task/cache-dependent list-rate estimates, not invoices.
Post-hoc review of 2,607 saved model-visible inputs flags 483/550 inputs containing extra calls (67 missed); this measures neither sensitivity across all updated-tool inputs nor online blocking.

\subsubsection{Same-byte environment comparison}\label{sec:environment-supplement}
For each frozen readable/protected package, five audit preparations and five ordinary-environment replays match the same byte map, with identical replay arguments (Table~\ref{tab:environment-components}).
Each package also receives one separate static review in the existing full-source configuration, without retries or changes to packages or completed runs.

\begin{table}[H]
\centering
\fontsize{7.5}{8.5}\selectfont
\renewcommand{\arraystretch}{0.88}
\caption{Separate outcomes from static review, runtime auditing, and ordinary replay of each frozen package. Each form has one static judgment and five runs per execution context.}
\label{tab:environment-components}
\setlength{\tabcolsep}{2.5pt}
\setlength{\aboverulesep}{0.6pt}
\setlength{\belowrulesep}{0.6pt}
\setlength{\cmidrulesep}{0.8pt}
\begin{tabular*}{.96\columnwidth}{@{\extracolsep{\fill}}lcccc@{}}
\toprule
Code form & Static & Audit allow & Audit effect & Replay effect \\
\midrule
Raw & BLOCK & 5/5 & 0/5 & 5/5 \\
Protected & ALLOW & 5/5 & 0/5 & 5/5 \\
\bottomrule
\end{tabular*}
\end{table}


%% file: main.bbl

\begin{thebibliography}{41}


\ifx \showCODEN    \undefined \def \showCODEN     #1{\unskip}     \fi
\ifx \showISBNx    \undefined \def \showISBNx     #1{\unskip}     \fi
\ifx \showISBNxiii \undefined \def \showISBNxiii  #1{\unskip}     \fi
\ifx \showISSN     \undefined \def \showISSN      #1{\unskip}     \fi
\ifx \showLCCN     \undefined \def \showLCCN      #1{\unskip}     \fi
\ifx \shownote     \undefined \def \shownote      #1{#1}          \fi
\ifx \showarticletitle \undefined \def \showarticletitle #1{#1}   \fi
\ifx \showURL      \undefined \def \showURL       {\relax}        \fi
\providecommand\bibfield[2]{#2}
\providecommand\bibinfo[2]{#2}
\providecommand\natexlab[1]{#1}
\providecommand\showeprint[2][]{arXiv:#2}
\makeatletter
\@ifundefined{NAT@parse@date}{}{\let\NAT@parse@date@orig\NAT@parse@date}
\@ifundefined{NAT@parse@date}{}{\def\NAT@parse@date#1#2#3#4#5#6@@{\NAT@parse@date@orig#1#2#3#4#5#6@@\def\NAT@tempyear{0000}\def\NAT@tempexlab{{?}}\ifx\NAT@year\NAT@tempyear\ifx\NAT@exlab\NAT@tempexlab\def\NAT@date{[n.\,d.]}\else\edef\NAT@date{[n.\,d.]\NAT@exlab}\fi\fi}}
\makeatother

\bibitem[{AWS Labs}(2026)]%
        {awsmcp2026}
\bibfield{author}{\bibinfo{person}{{AWS Labs}}.}
  \bibinfo{year}{2026}\natexlab{}.
\newblock \bibinfo{title}{{MCP Servers for AWS}}.
\newblock \bibinfo{howpublished}{Software repository}.
\newblock
\urldef\tempurl%
\url{https://github.com/awslabs/mcp/tree/d534d0fa3c06838336ac1ed64e7114db0ba85e6f}
\showURL{%
\tempurl}
\newblock
\shownote{Snapshot used for the historical-update sample}.


\bibitem[Beurer-Kellner and Fischer(2025)]%
        {invariant2025rugpull}
\bibfield{author}{\bibinfo{person}{Luca Beurer-Kellner} {and}
  \bibinfo{person}{Marc Fischer}.} \bibinfo{year}{2025}\natexlab{}.
\newblock \bibinfo{title}{{MCP} Security Notification: Tool Poisoning Attacks}.
\newblock \bibinfo{howpublished}{Invariant Labs Blog,
  \url{https://invariantlabs.ai/blog/mcp-security-notification-tool-poisoning-attacks}}.
\newblock
\newblock
\shownote{Accessed: 2026-10-17}.


\bibitem[Bhatt et~al\mbox{.}(2025)]%
        {etdi2025}
\bibfield{author}{\bibinfo{person}{Manish Bhatt}, \bibinfo{person}{Vineeth~Sai
  Narajala}, {and} \bibinfo{person}{Idan Habler}.}
  \bibinfo{year}{2025}\natexlab{}.
\newblock \showarticletitle{{ETDI}: Mitigating Tool Squatting and Rug Pull
  Attacks in Model Context Protocol ({MCP}) by Using {OAuth}-Enhanced Tool
  Definitions and Policy-Based Access Control}.
\newblock \bibinfo{journal}{\emph{arXiv preprint arXiv:2506.01333}}
  (\bibinfo{year}{2025}).
\newblock


\bibitem[{bubblewrap contributors}(0000)]%
        {bubblewrap}
\bibfield{author}{\bibinfo{person}{{bubblewrap contributors}}.}
  \bibinfo{year}{[n.\,d.]}\natexlab{}.
\newblock \bibinfo{title}{bubblewrap: A Low-Level Unprivileged Sandboxing
  Tool}.
\newblock \bibinfo{howpublished}{Software repository}.
\newblock
\urldef\tempurl%
\url{https://github.com/containers/bubblewrap}
\showURL{%
\tempurl}
\newblock
\shownote{Accessed: October 25, 2026}.


\bibitem[Chang et~al\mbox{.}(2026)]%
        {ipi-chang}
\bibfield{author}{\bibinfo{person}{Hongyan Chang}, \bibinfo{person}{Ergute
  Bao}, \bibinfo{person}{Xinjian Luo}, {and} \bibinfo{person}{Ting Yu}.}
  \bibinfo{year}{2026}\natexlab{}.
\newblock \showarticletitle{Overcoming the Retrieval Barrier: Indirect Prompt
  Injection in the Wild for {LLM} Systems}.
\newblock \bibinfo{journal}{\emph{CoRR}}  \bibinfo{volume}{abs/2601.07072}
  (\bibinfo{year}{2026}).
\newblock


\bibitem[{Cisco AI Defense}(2026)]%
        {ciscomcpscanner2026}
\bibfield{author}{\bibinfo{person}{{Cisco AI Defense}}.}
  \bibinfo{year}{2026}\natexlab{}.
\newblock \bibinfo{title}{{MCP Scanner}}.
\newblock \bibinfo{howpublished}{Software repository}.
\newblock
\urldef\tempurl%
\url{https://github.com/cisco-ai-defense/mcp-scanner}
\showURL{%
\tempurl}
\newblock
\shownote{Accessed 2026-10-24}.


\bibitem[Debenedetti et~al\mbox{.}(2025)]%
        {camel2025}
\bibfield{author}{\bibinfo{person}{Edoardo Debenedetti}, \bibinfo{person}{Ilia
  Shumailov}, \bibinfo{person}{Tianqi Fan}, \bibinfo{person}{Jamie Hayes},
  \bibinfo{person}{Nicholas Carlini}, \bibinfo{person}{Daniel Fabian},
  \bibinfo{person}{Christoph Kern}, \bibinfo{person}{Chongyang Shi},
  \bibinfo{person}{Andreas Terzis}, {and} \bibinfo{person}{Florian
  Tram{\`e}r}.} \bibinfo{year}{2025}\natexlab{}.
\newblock \showarticletitle{Defeating Prompt Injections by Design}.
\newblock \bibinfo{journal}{\emph{arXiv preprint arXiv:2503.18813}}
  (\bibinfo{year}{2025}).
\newblock
\urldef\tempurl%
\url{https://arxiv.org/abs/2503.18813}
\showURL{%
\tempurl}
\newblock
\shownote{Version 2}.


\bibitem[Debenedetti et~al\mbox{.}(2024)]%
        {agentdojo2024}
\bibfield{author}{\bibinfo{person}{Edoardo Debenedetti}, \bibinfo{person}{Jie
  Zhang}, \bibinfo{person}{Mislav Balunovi{\'c}}, \bibinfo{person}{Luca
  Beurer-Kellner}, \bibinfo{person}{Marc Fischer}, {and}
  \bibinfo{person}{Florian Tram{\`e}r}.} \bibinfo{year}{2024}\natexlab{}.
\newblock \showarticletitle{{AgentDojo}: A Dynamic Environment to Evaluate
  Prompt Injection Attacks and Defenses for {LLM} Agents}. In
  \bibinfo{booktitle}{\emph{Advances in Neural Information Processing
  Systems}}, Vol.~\bibinfo{volume}{37}. \bibinfo{pages}{82895--82920}.
\newblock
\href{https://doi.org/10.52202/079017-2636}{doi:\nolinkurl{10.52202/079017-2636}}


\bibitem[Dragos et~al\mbox{.}(2024)]%
        {xzutils2024}
\bibfield{author}{\bibinfo{person}{Dumitra Dragos}, \bibinfo{person}{Amit
  Serper}, {and} \bibinfo{person}{Suraj Sahu}.}
  \bibinfo{year}{2024}\natexlab{}.
\newblock \bibinfo{title}{{CVE}-2024-3094 and the {XZ} Upstream Supply Chain
  Attack: What You Need to Know}.
\newblock \bibinfo{howpublished}{CrowdStrike Blog,
  \url{https://www.crowdstrike.com/en-us/blog/cve-2024-3094-xz-upstream-supply-chain-attack/}}.
\newblock
\newblock
\shownote{Accessed: 2026-09-23}.


\bibitem[{GitHub}(2026a)]%
        {githubmcp2026}
\bibfield{author}{\bibinfo{person}{{GitHub}}.}
  \bibinfo{year}{2026}\natexlab{a}.
\newblock \bibinfo{title}{{GitHub MCP Server}}.
\newblock \bibinfo{howpublished}{Software repository}.
\newblock
\urldef\tempurl%
\url{https://github.com/github/github-mcp-server/tree/e3809fe91c7546633241006bf0b91f23b2595e28}
\showURL{%
\tempurl}
\newblock
\shownote{Snapshot used for the historical-update sample}.


\bibitem[{GitHub}(2026b)]%
        {githubmcp2032026}
\bibfield{author}{\bibinfo{person}{{GitHub}}.}
  \bibinfo{year}{2026}\natexlab{b}.
\newblock \bibinfo{title}{{GitHub MCP Server v2.0.3}}.
\newblock \bibinfo{howpublished}{Software release}.
\newblock
\urldef\tempurl%
\url{https://github.com/github/github-mcp-server/releases/tag/v2.0.3}
\showURL{%
\tempurl}


\bibitem[Greshake et~al\mbox{.}(2023)]%
        {greshake2023}
\bibfield{author}{\bibinfo{person}{Kai Greshake}, \bibinfo{person}{Sahar
  Abdelnabi}, \bibinfo{person}{Shailesh Mishra}, \bibinfo{person}{Christoph
  Endres}, \bibinfo{person}{Thorsten Holz}, {and} \bibinfo{person}{Mario
  Fritz}.} \bibinfo{year}{2023}\natexlab{}.
\newblock \showarticletitle{Not What You've Signed Up For: Compromising
  Real-World {LLM}-Integrated Applications with Indirect Prompt Injection}.
\newblock \bibinfo{journal}{\emph{arXiv preprint arXiv:2302.12173}}
  (\bibinfo{year}{2023}).
\newblock


\bibitem[Guo et~al\mbox{.}(2026)]%
        {intelguard2026}
\bibfield{author}{\bibinfo{person}{Wenbo Guo}, \bibinfo{person}{Shiwen Song},
  \bibinfo{person}{Jiaxun Guo}, \bibinfo{person}{Zhengzi Xu},
  \bibinfo{person}{Chengwei Liu}, \bibinfo{person}{Haoran Ou},
  \bibinfo{person}{Mengmeng Ge}, {and} \bibinfo{person}{Yang Liu}.}
  \bibinfo{year}{2026}\natexlab{}.
\newblock \showarticletitle{Bridging Expert Reasoning and LLM Detection: A
  Knowledge-Driven Framework for Malicious Packages}. In
  \bibinfo{booktitle}{\emph{Proceedings of the ACM Web Conference 2026}}.
\newblock
\href{https://doi.org/10.1145/3774904.3792083}{doi:\nolinkurl{10.1145/3774904.3792083}}


\bibitem[Hu et~al\mbox{.}(2026)]%
        {maltool2026}
\bibfield{author}{\bibinfo{person}{Yuepeng Hu}, \bibinfo{person}{Yuqi Jia},
  \bibinfo{person}{Mengyuan Li}, \bibinfo{person}{Dawn Song}, {and}
  \bibinfo{person}{Neil Gong}.} \bibinfo{year}{2026}\natexlab{}.
\newblock \showarticletitle{{MalTool}: Malicious Tool Attacks on {LLM} Agents}.
\newblock \bibinfo{journal}{\emph{arXiv preprint arXiv:2602.12194}}
  (\bibinfo{year}{2026}).
\newblock


\bibitem[Huang et~al\mbox{.}(2026)]%
        {mcpthreatmodel2026}
\bibfield{author}{\bibinfo{person}{Charoes Huang}, \bibinfo{person}{Xin Huang},
  \bibinfo{person}{Ngoc~Phu Tran}, {and} \bibinfo{person}{Amin~Milani Fard}.}
  \bibinfo{year}{2026}\natexlab{}.
\newblock \showarticletitle{Model Context Protocol Threat Modeling and
  Analyzing Vulnerabilities to Prompt Injection with Tool Poisoning}.
\newblock \bibinfo{journal}{\emph{arXiv preprint arXiv:2603.22489}}
  (\bibinfo{year}{2026}).
\newblock


\bibitem[Ji et~al\mbox{.}(2026)]%
        {cloakdetonate2026}
\bibfield{author}{\bibinfo{person}{Zimo Ji}, \bibinfo{person}{Congying Xu},
  \bibinfo{person}{Zongjie Li}, \bibinfo{person}{Yudong Gao},
  \bibinfo{person}{Xin Wei}, \bibinfo{person}{Shuai Wang}, {and}
  \bibinfo{person}{Shing-Chi Cheung}.} \bibinfo{year}{2026}\natexlab{}.
\newblock \showarticletitle{Cloak and Detonate: Scanner Evasion and Dynamic
  Detection of Agent Skill Malware}.
\newblock \bibinfo{journal}{\emph{arXiv preprint arXiv:2607.02357}}
  (\bibinfo{year}{2026}).
\newblock
\urldef\tempurl%
\url{https://arxiv.org/abs/2607.02357}
\showURL{%
\tempurl}


\bibitem[Li et~al\mbox{.}(2026)]%
        {deepagent2026}
\bibfield{author}{\bibinfo{person}{Xiaoxi Li}, \bibinfo{person}{Wenxiang Jiao},
  \bibinfo{person}{Jiarui Jin}, \bibinfo{person}{Guanting Dong},
  \bibinfo{person}{Jiajie Jin}, \bibinfo{person}{Yinuo Wang},
  \bibinfo{person}{Hao Wang}, \bibinfo{person}{Yutao Zhu},
  \bibinfo{person}{Ji-Rong Wen}, \bibinfo{person}{Yuan Lu}, {and}
  \bibinfo{person}{Zhicheng Dou}.} \bibinfo{year}{2026}\natexlab{}.
\newblock \showarticletitle{{DeepAgent}: A General Reasoning Agent with
  Scalable Toolsets}. In \bibinfo{booktitle}{\emph{Proceedings of the ACM Web
  Conference 2026}}.
\newblock
\href{https://doi.org/10.1145/3774904.3792460}{doi:\nolinkurl{10.1145/3774904.3792460}}


\bibitem[Liu et~al\mbox{.}(2026b)]%
        {payloadless2026}
\bibfield{author}{\bibinfo{person}{Xinyu Liu}, \bibinfo{person}{Xing Hu},
  \bibinfo{person}{Yukai Zhao}, {and} \bibinfo{person}{Xin Xia}.}
  \bibinfo{year}{2026}\natexlab{b}.
\newblock \showarticletitle{Exploiting {LLM} Agent Supply Chains via
  Payload-less Skills}.
\newblock \bibinfo{journal}{\emph{arXiv preprint arXiv:2605.14460}}
  (\bibinfo{year}{2026}).
\newblock


\bibitem[Liu et~al\mbox{.}(2026a)]%
        {maliciousskillswild2026}
\bibfield{author}{\bibinfo{person}{Yi Liu}, \bibinfo{person}{Zhihao Chen},
  \bibinfo{person}{Yanjun Zhang}, \bibinfo{person}{Gelei Deng},
  \bibinfo{person}{Yuekang Li}, \bibinfo{person}{Jianting Ning}, {and}
  \bibinfo{person}{Leo~Yu Zhang}.} \bibinfo{year}{2026}\natexlab{a}.
\newblock \showarticletitle{"Do Not Mention This to the User": Detecting and
  Understanding Malicious Agent Skills in the Wild}. In
  \bibinfo{booktitle}{\emph{35th USENIX Security Symposium (USENIX Security
  26)}}. \bibinfo{pages}{1727--1746}.
\newblock
\urldef\tempurl%
\url{https://www.usenix.org/conference/usenixsecurity26/presentation/liu-yi}
\showURL{%
\tempurl}


\bibitem[{Model Context Protocol}(2026)]%
        {mcpgit2026}
\bibfield{author}{\bibinfo{person}{{Model Context Protocol}}.}
  \bibinfo{year}{2026}\natexlab{}.
\newblock \bibinfo{title}{{Model Context Protocol Servers}}.
\newblock \bibinfo{howpublished}{Software repository}.
\newblock
\urldef\tempurl%
\url{https://github.com/modelcontextprotocol/servers/tree/35becbdacc22881347d08833a87dc9e6ea3dfedd}
\showURL{%
\tempurl}
\newblock
\shownote{Source snapshot used for the Python supplementary instance}.


\bibitem[Moser et~al\mbox{.}(2007)]%
        {4413008}
\bibfield{author}{\bibinfo{person}{Andreas Moser}, \bibinfo{person}{Christopher
  Kruegel}, {and} \bibinfo{person}{Engin Kirda}.}
  \bibinfo{year}{2007}\natexlab{}.
\newblock \showarticletitle{Limits of Static Analysis for Malware Detection}.
  In \bibinfo{booktitle}{\emph{Twenty-Third Annual Computer Security
  Applications Conference (ACSAC 2007)}}. \bibinfo{pages}{421--430}.
\newblock
\href{https://doi.org/10.1109/ACSAC.2007.21}{doi:\nolinkurl{10.1109/ACSAC.2007.21}}


\bibitem[{npm, Inc.}(2018)]%
        {eventstream2018}
\bibfield{author}{\bibinfo{person}{{npm, Inc.}}}
  \bibinfo{year}{2018}\natexlab{}.
\newblock \bibinfo{title}{Details about the event-stream incident}.
\newblock \bibinfo{howpublished}{The npm Blog,
  \url{https://blog.npmjs.org/post/180565383195/details-about-the-event-stream-incident}}.
\newblock
\newblock
\shownote{Accessed: 2026-09-23}.


\bibitem[{NVIDIA}(2026)]%
        {skillspector2026}
\bibfield{author}{\bibinfo{person}{{NVIDIA}}.} \bibinfo{year}{2026}\natexlab{}.
\newblock \bibinfo{title}{{SkillSpector}}.
\newblock \bibinfo{howpublished}{\url{https://github.com/NVIDIA/SkillSpector}}.
\newblock
\newblock
\shownote{Version 2.12.0, commit 11c8b48e35056602fad263438448e0ffb5be2776.
  Accessed October 21, 2026}.


\bibitem[Oliver et~al\mbox{.}(2013)]%
        {oliver2013tlsh}
\bibfield{author}{\bibinfo{person}{Jonathan Oliver}, \bibinfo{person}{Chun
  Cheng}, {and} \bibinfo{person}{Yanggui Chen}.}
  \bibinfo{year}{2013}\natexlab{}.
\newblock \showarticletitle{{TLSH} -- A Locality Sensitive Hash}. In
  \bibinfo{booktitle}{\emph{2013 Fourth Cybercrime and Trustworthy Computing
  Workshop}}. \bibinfo{publisher}{IEEE}, \bibinfo{pages}{7--13}.
\newblock
\href{https://doi.org/10.1109/CTC.2013.9}{doi:\nolinkurl{10.1109/CTC.2013.9}}


\bibitem[Qu et~al\mbox{.}(2026)]%
        {skillpoisoning2026}
\bibfield{author}{\bibinfo{person}{Yubin Qu}, \bibinfo{person}{Yi Liu},
  \bibinfo{person}{Tongcheng Geng}, \bibinfo{person}{Gelei Deng},
  \bibinfo{person}{Yuekang Li}, \bibinfo{person}{Leo~Yu Zhang},
  \bibinfo{person}{Ying Zhang}, {and} \bibinfo{person}{Lei Ma}.}
  \bibinfo{year}{2026}\natexlab{}.
\newblock \showarticletitle{Supply-Chain Poisoning Attacks Against {LLM} Coding
  Agent Skill Ecosystems}.
\newblock \bibinfo{journal}{\emph{arXiv preprint arXiv:2604.03081}}
  (\bibinfo{year}{2026}).
\newblock


\bibitem[{Qualixar}(2026)]%
        {skillfortify2026}
\bibfield{author}{\bibinfo{person}{{Qualixar}}.}
  \bibinfo{year}{2026}\natexlab{}.
\newblock \bibinfo{title}{{SkillFortify}}.
\newblock
  \bibinfo{howpublished}{\url{https://github.com/qualixar/skillfortify}}.
\newblock
\newblock
\shownote{Version 0.6.0, commit dbb5942deae46afdb5b85440d5f5d851c197c30e.
  Accessed October 21, 2026}.


\bibitem[Rao et~al\mbox{.}(2026)]%
        {acoda2026}
\bibfield{author}{\bibinfo{person}{Hongzhou Rao}, \bibinfo{person}{Zikan Dong},
  \bibinfo{person}{Yanjie Zhao}, \bibinfo{person}{Haodong Li}, {and}
  \bibinfo{person}{Haoyu Wang}.} \bibinfo{year}{2026}\natexlab{}.
\newblock \showarticletitle{Acoda: Adversarial Code Obfuscation for Defending
  against {LLM}-based Analysis}. In \bibinfo{booktitle}{\emph{Companion
  Proceedings of the 48th IEEE/ACM International Conference on Software
  Engineering (ICSE-Companion), Industry Challenge Track}}.
\newblock
\newblock
\shownote{arXiv:2606.11755}.


\bibitem[Schick et~al\mbox{.}(2023)]%
        {schick2023toolformer}
\bibfield{author}{\bibinfo{person}{Timo Schick}, \bibinfo{person}{Jane
  Dwivedi-Yu}, \bibinfo{person}{Roberto Dess{\`i}}, \bibinfo{person}{Roberta
  Raileanu}, \bibinfo{person}{Maria Lomeli}, \bibinfo{person}{Eric Hambro},
  \bibinfo{person}{Luke Zettlemoyer}, \bibinfo{person}{Nicola Cancedda}, {and}
  \bibinfo{person}{Thomas Scialom}.} \bibinfo{year}{2023}\natexlab{}.
\newblock \showarticletitle{Toolformer: Language Models Can Teach Themselves to
  Use Tools}. In \bibinfo{booktitle}{\emph{Advances in Neural Information
  Processing Systems (NeurIPS)}}.
\newblock


\bibitem[Shi et~al\mbox{.}(2026)]%
        {toolhijacker2026}
\bibfield{author}{\bibinfo{person}{Jiawen Shi}, \bibinfo{person}{Zenghui Yuan},
  \bibinfo{person}{Guiyao Tie}, \bibinfo{person}{Pan Zhou},
  \bibinfo{person}{Neil~Zhenqiang Gong}, {and} \bibinfo{person}{Lichao Sun}.}
  \bibinfo{year}{2026}\natexlab{}.
\newblock \showarticletitle{Prompt Injection Attack to Tool Selection in {LLM}
  Agents}. In \bibinfo{booktitle}{\emph{Network and Distributed System Security
  Symposium (NDSS)}}.
\newblock
\urldef\tempurl%
\url{https://www.ndss-symposium.org/ndss-paper/prompt-injection-attack-to-tool-selection-in-llm-agents/}
\showURL{%
\tempurl}


\bibitem[{Sigstore}(2026)]%
        {sigstoreverify2026}
\bibfield{author}{\bibinfo{person}{{Sigstore}}.}
  \bibinfo{year}{2026}\natexlab{}.
\newblock \bibinfo{title}{Verifying Signatures}.
\newblock
  \bibinfo{howpublished}{\url{https://docs.sigstore.dev/cosign/verifying/verify/}}.
\newblock
\newblock
\shownote{Accessed October 21, 2026}.


\bibitem[{SLSA}(2026)]%
        {slsaverify2026}
\bibfield{author}{\bibinfo{person}{{SLSA}}.} \bibinfo{year}{2026}\natexlab{}.
\newblock \bibinfo{title}{Build: Verifying Artifacts}.
\newblock
  \bibinfo{howpublished}{\url{https://slsa.dev/spec/v1.2/verifying-artifacts}}.
\newblock
\newblock
\shownote{Version 1.2; accessed October 21, 2026}.


\bibitem[Stein(2026)]%
        {mcpscale2026}
\bibfield{author}{\bibinfo{person}{Merlin Stein}.}
  \bibinfo{year}{2026}\natexlab{}.
\newblock \showarticletitle{How Are {AI} Agents Used? {E}vidence from 177,000
  {MCP} Tools}.
\newblock \bibinfo{journal}{\emph{arXiv preprint arXiv:2603.23802}}
  (\bibinfo{year}{2026}).
\newblock


\bibitem[Wang et~al\mbox{.}(2026a)]%
        {malskills2026}
\bibfield{author}{\bibinfo{person}{Shenao Wang}, \bibinfo{person}{Junjie He},
  \bibinfo{person}{Yanjie Zhao}, \bibinfo{person}{Yayi Wang},
  \bibinfo{person}{Kan Yu}, {and} \bibinfo{person}{Haoyu Wang}.}
  \bibinfo{year}{2026}\natexlab{a}.
\newblock \showarticletitle{MalSkills: Detecting Malicious Skills in the
  Agentic Supply Chain via Neuro-symbolic Reasoning}.
\newblock \bibinfo{journal}{\emph{arXiv preprint arXiv:2603.27204}}
  (\bibinfo{year}{2026}).
\newblock
\urldef\tempurl%
\url{https://arxiv.org/abs/2603.27204}
\showURL{%
\tempurl}


\bibitem[Wang et~al\mbox{.}(2025a)]%
        {mcptox2025}
\bibfield{author}{\bibinfo{person}{Zhiqiang Wang}, \bibinfo{person}{Yichao
  Gao}, \bibinfo{person}{Yanting Wang}, \bibinfo{person}{Suyuan Liu},
  \bibinfo{person}{Haifeng Sun}, \bibinfo{person}{Haoran Cheng},
  \bibinfo{person}{Guanquan Shi}, \bibinfo{person}{Haohua Du}, {and}
  \bibinfo{person}{Xiangyang Li}.} \bibinfo{year}{2025}\natexlab{a}.
\newblock \showarticletitle{{MCPTox}: A Benchmark for Tool Poisoning Attack on
  Real-World {MCP} Servers}.
\newblock \bibinfo{journal}{\emph{arXiv preprint arXiv:2508.14925}}
  (\bibinfo{year}{2025}).
\newblock


\bibitem[Wang et~al\mbox{.}(2026b)]%
        {mcpfixgen2026}
\bibfield{author}{\bibinfo{person}{Zhiqiang Wang}, \bibinfo{person}{Guanquan
  Shi}, \bibinfo{person}{Yanting Wang}, \bibinfo{person}{Yichao Gao},
  \bibinfo{person}{Hongsen Lang}, \bibinfo{person}{Yunhao Yao},
  \bibinfo{person}{Haohua Du}, {and} \bibinfo{person}{Xiang-Yang Li}.}
  \bibinfo{year}{2026}\natexlab{b}.
\newblock \showarticletitle{Beyond Detection: Autonomous Anomaly Remediation
  for {MCP} Against Tool Poisoning Attacks}. In
  \bibinfo{booktitle}{\emph{Proceedings of the ACM Web Conference 2026}}.
  \bibinfo{pages}{2974--2982}.
\newblock
\href{https://doi.org/10.1145/3774904.3792400}{doi:\nolinkurl{10.1145/3774904.3792400}}


\bibitem[Wang et~al\mbox{.}(2025b)]%
        {agentvigil2025}
\bibfield{author}{\bibinfo{person}{Zhun Wang}, \bibinfo{person}{Vincent Siu},
  \bibinfo{person}{Zhe Ye}, \bibinfo{person}{Tianneng Shi},
  \bibinfo{person}{Yuzhou Nie}, \bibinfo{person}{Xuandong Zhao},
  \bibinfo{person}{Chenguang Wang}, \bibinfo{person}{Wenbo Guo}, {and}
  \bibinfo{person}{Dawn Song}.} \bibinfo{year}{2025}\natexlab{b}.
\newblock \showarticletitle{{AGENTVIGIL}: Automatic Black-Box Red-teaming for
  Indirect Prompt Injection against {LLM} Agents}. In
  \bibinfo{booktitle}{\emph{Findings of the Association for Computational
  Linguistics: EMNLP 2025}}. \bibinfo{publisher}{Association for Computational
  Linguistics}, \bibinfo{address}{Suzhou, China},
  \bibinfo{pages}{23159--23172}.
\newblock
\href{https://doi.org/10.18653/v1/2025.findings-emnlp.1258}{doi:\nolinkurl{10.18653/v1/2025.findings-emnlp.1258}}


\bibitem[Yao et~al\mbox{.}(2023)]%
        {yao2023react}
\bibfield{author}{\bibinfo{person}{Shunyu Yao}, \bibinfo{person}{Jeffrey Zhao},
  \bibinfo{person}{Dian Yu}, \bibinfo{person}{Nan Du}, \bibinfo{person}{Izhak
  Shafran}, \bibinfo{person}{Karthik Narasimhan}, {and} \bibinfo{person}{Yuan
  Cao}.} \bibinfo{year}{2023}\natexlab{}.
\newblock \showarticletitle{{ReAct}: Synergizing Reasoning and Acting in
  Language Models}. In \bibinfo{booktitle}{\emph{Proceedings of the
  International Conference on Learning Representations (ICLR)}}.
\newblock


\bibitem[Yi et~al\mbox{.}(2025)]%
        {bipia2025}
\bibfield{author}{\bibinfo{person}{Jingwei Yi}, \bibinfo{person}{Yueqi Xie},
  \bibinfo{person}{Bin Zhu}, \bibinfo{person}{Emre Kiciman},
  \bibinfo{person}{Guangzhong Sun}, \bibinfo{person}{Xing Xie}, {and}
  \bibinfo{person}{Fangzhao Wu}.} \bibinfo{year}{2025}\natexlab{}.
\newblock \showarticletitle{Benchmarking and Defending against Indirect Prompt
  Injection Attacks on Large Language Models}. In
  \bibinfo{booktitle}{\emph{Proceedings of the 31st ACM SIGKDD Conference on
  Knowledge Discovery and Data Mining}}.
\newblock
\href{https://doi.org/10.1145/3690624.3709179}{doi:\nolinkurl{10.1145/3690624.3709179}}


\bibitem[Yokoyama et~al\mbox{.}(2016)]%
        {yokoyama2016sandprint}
\bibfield{author}{\bibinfo{person}{Akira Yokoyama}, \bibinfo{person}{Kou
  Ishii}, \bibinfo{person}{Rui Tanabe}, \bibinfo{person}{Yinmin Papa},
  \bibinfo{person}{Katsunari Yoshioka}, \bibinfo{person}{Tsutomu Matsumoto},
  \bibinfo{person}{Takahiro Kasama}, \bibinfo{person}{Daisuke Inoue},
  \bibinfo{person}{Michael Brengel}, \bibinfo{person}{Michael Backes}, {and}
  \bibinfo{person}{Christian Rossow}.} \bibinfo{year}{2016}\natexlab{}.
\newblock \showarticletitle{{SandPrint}: Fingerprinting Malware Sandboxes to
  Provide Intelligence for Sandbox Evasion}. In
  \bibinfo{booktitle}{\emph{Proceedings of the 19th International Symposium on
  Research in Attacks, Intrusions, and Defenses (RAID)}}.
\newblock


\bibitem[Zhan et~al\mbox{.}(2024)]%
        {injecagent2024}
\bibfield{author}{\bibinfo{person}{Qiusi Zhan}, \bibinfo{person}{Zhixiang
  Liang}, \bibinfo{person}{Zifan Ying}, {and} \bibinfo{person}{Daniel Kang}.}
  \bibinfo{year}{2024}\natexlab{}.
\newblock \showarticletitle{{InjecAgent}: Benchmarking Indirect Prompt
  Injections in Tool-Integrated Large Language Model Agents}. In
  \bibinfo{booktitle}{\emph{Findings of the Association for Computational
  Linguistics: ACL 2024}}. \bibinfo{pages}{10471--10506}.
\newblock
\href{https://doi.org/10.18653/v1/2024.findings-acl.624}{doi:\nolinkurl{10.18653/v1/2024.findings-acl.624}}


\bibitem[Zhao et~al\mbox{.}(2026)]%
        {parasites2026}
\bibfield{author}{\bibinfo{person}{Shuli Zhao}, \bibinfo{person}{Qinsheng Hou},
  \bibinfo{person}{Zihan Zhan}, \bibinfo{person}{Yanhao Wang},
  \bibinfo{person}{Yuchong Xie}, \bibinfo{person}{Yu Guo},
  \bibinfo{person}{Libo Chen}, \bibinfo{person}{Shenghong Li}, {and}
  \bibinfo{person}{Zhi Xue}.} \bibinfo{year}{2026}\natexlab{}.
\newblock \showarticletitle{Parasites in the Toolchain: A Large-Scale Analysis
  of Attacks on the {MCP} Ecosystem}. In \bibinfo{booktitle}{\emph{2026 IEEE
  Symposium on Security and Privacy (SP)}}. \bibinfo{publisher}{IEEE},
  \bibinfo{pages}{138--155}.
\newblock
\href{https://doi.org/10.1109/SP63933.2026.00154}{doi:\nolinkurl{10.1109/SP63933.2026.00154}}


\end{thebibliography}
